\documentclass[12pt, reqno]{amsart}

\usepackage{amssymb} 

\usepackage{graphicx}

\newcommand{\wt}{\widetilde}

\newtheorem{dfn}{Definition}
\newtheorem{prop}{Proposition}
\newtheorem{thm}{Theorem}

\newcommand{\wh}[1]{\widehat{#1}}

\begin{document}

\title[Pauli problem in thermodynamics]{Pauli problem in thermodynamics}

\author[A. E. Ruuge]{Artur E. Ruuge}
\address{
Department of Mathematics and Computer Science, 
University of Antwerp, 
Middelheim Campus Building G, 
Middelheimlaan 1, B-2020, 
Antwerp, Belgium
}
\email{artur.ruuge@ua.ac.be}

\begin{abstract}
A thermodynamic analogue of the Pauli problem 
(reconstruction of a wavefunction from the position and momentum distributions) is formulated. 
The coordinates of a quantum system are replaced by the inverse absolute temperature and other intensive quantities, 
and the Planck constant is replaced by the Boltzmann constant multiplied by two. 
A new natural mathematical generalization of the quasithermodynamic fluctuation theory is suggested and 
sufficient conditions for the existence of asymptotic solutions of the thermodynamic Pauli problem are obtained. 
\end{abstract}

\maketitle

\section{Introduction}

Anyone who had studied statistical thermodynamics could have noticed a certain analogy between the 
mathematical formulae describing the quasithermodynamic fluctuations and the 
quantum mechanical formulae related related to coherent states and the Heisenberg uncertainty principle. 
Take a simple example of an abstract thermodynamic system with an entropy function 
\begin{equation*} 
S = S (U, V, N), 
\end{equation*}
where $U$ is internal energy, $V$ is volume, and $N$ is the number of moles of a chemical substance 
(one may think of a Van der Waals gas).
If we fix the volume $V = V_0$ and the number of moles $N = N_0$, and 
put the system in a thermostat with absolute temperature $T = T_0$, 
then the internal energy $U$ is going to experience fluctuations $\delta U$ 
around a value $U = U_0$ satisfying an equation $T_{0}^{-1} = \partial S (U, V_0, N_0)/ \partial U$ 
(we assume this solution is unique). 
The corresponding probability density $f_{\delta U} (x)$ in a point $x \in \mathbb{R}$ is approximately described by the 
normal distribution (Einstein's formula): 
\begin{equation*} 
f_{\delta U} (x) = (2 \pi k_B [- S_{U, U}'']^{-1})^{- 1/ 2} \exp ( S_{U, U}'' x^2/ (2 k_B) ), 
\end{equation*}
where $k_B$ is the Boltzmann constant, and 
$S_{U, U}''$ is a short notation for the second derivative 
$\partial^2 S (U, V, N)/ \partial U^2$ taken in $U = U_0$, $V = V_0$, $N = N_0$ 
(we assume that $S_{U, U}'' < 0$). 
The variance of $\delta U$ is then as follows: 
\begin{equation*} 
\langle (\delta U)^2 \rangle = k_B [- S_{U, U}'']^{-1}. 
\end{equation*}
It is quite important to stress that 
the internal energy $U$ is not the only quantity that is going to experience fluctuations. 
We have at the same time a fluctuation $\delta \beta$ of 
the inverse absolute temperature $\beta = \partial S (U, V_0, N_0)/ \partial U$ around the value $\beta_0 := 1/ T_0$. 
Its probability density $f_{\delta \beta} (y)$ in a point $y \in \mathbb{R}$ is approximated by the formula: 
\begin{equation*} 
f_{\delta \beta} (y) = (2 \pi k_B [- S_{U, U}''])^{- 1/ 2} \exp ( y^2/ (2 k_B S_{U, U}'') ).  
\end{equation*}
Note that $f_{\delta \beta} (y)$ coincides with the probability density of the linearly transformed random variable $S_{U, U}'' \delta U$. 
For the variance of $\delta \beta$ we have: 
\begin{equation*} 
\langle (\delta \beta)^{2} \rangle = k_B [- S_{U, U}''], 
\end{equation*}
so we obtain: 
\begin{equation*} 
\sqrt{\langle (\delta U)^2 \rangle} 
\sqrt{\langle (\delta \beta)^2 \rangle} 
= k_B.
\end{equation*}

All this is very similar to the Heisenberg uncertainty relation in a coherent state. 
This observation was made by many authors, see for example 
\cite{BalianValentin, Kazinski, Lavenda, Mehrafarin, RudoiSukhanov, UffinkLith, VelazquezCurilef}. 
In fact the discussion of this analogy can be traced back to W.~Heisenberg and N.~Bohr 
(the complementarity principle). 
The \emph{non-equilibrium} thermodynamics 
is a less explored area, but 
one can point out that certain stochastic models describing weakly non-equilibrium states  
admit a description in terms of Feynman path integrals  
\cite{OnsagerMachlup, MachlupOnsager, AcostaFernandez-de-CordobaIsidroSantander, Fernandez-de-CordobaIsidroPerea}.

Take an abstract one-dimensional quantum mechanical system 
described by a wavefunction 
\begin{equation*} 
\psi_{\hbar} (q) = \Big( \frac{a}{\pi \hbar} \Big)^{1/ 4} 
\exp \Big( - \frac{a q^2}{2 \hbar} \Big), 
\end{equation*}
where $a > 0$ is a parameter, $q$ varies over $\mathbb{R}$ 
corresponding to the possible values of a coordinate $Q$, 
$\hbar$ is the Planck constant, and $\psi_{\hbar} \in L^2 (\mathbb{R})$. 
The self-adjoint operator $\wh{Q}$ representing $Q$ acts as a multiplication by $q$, 
and the self-adjoint operator $\wh{P}$ representing the canonically conjugate momentum $P$ 
acts as a derivation $- i \hbar \partial/ \partial q$. 
The state corresponding to $\psi_{\hbar}$ is coherent, i.e. 
we reach the lower bound in the Heisenberg uncertainty relation:
\begin{equation*} 
\sqrt{\langle (\delta Q)^2 \rangle_{\hbar}} 
\sqrt{\langle (\delta P)^2 \rangle_{\hbar}}
= \hbar/ 2, 
\end{equation*}
where $\langle - \rangle_{\hbar}$ denotes the quantum mechanical average, 
$\delta Q = Q - \langle Q \rangle_{\hbar}$, and $\delta P = P - \langle P \rangle_{\hbar}$. 

The analogy between the two formulae is too striking to neglect. 
The Planck constant $\hbar$ corresponds to the Boltzmann constant $k_B$ multiplied by two, 
but to develop this analogy one needs answer the following question: 
What is a \emph{``thermodynamic wavefunction''}? 
In the present paper I suggest to attack this question via the so-called \emph{Pauli problem}. 
The Pauli problem is basically a problem of a \emph{reconstruction} of a quantum mechanical wave function 
based on the directly available experimental data. 
Around 1933 W. Pauli has asked the following: to what extent is a quantum mechanical 
wavefunction $\psi (q) \in L^2 (\mathbb{R})$ determined by the corresponding marginal distributions 
\begin{equation*} 
I (q) = |\psi (q)|^2, \quad 
J (p) = |\wt{\psi} (p)|^2, 
\end{equation*}
 of coordinates $q \in \mathbb{R}$ and momenta $p \in \mathbb{R}$, 
\begin{equation*} 
\wt{\psi} (p) := \frac{1}{(2 \pi \hbar)^{1/ 2}} \int_{\mathbb{R}} d q\, 
e^{i p q/ \hbar} \psi (q).
\end{equation*}
The probability densities $I (q)$ and $J (p)$ (assuming $\| \psi \| = 1$) is something we can 
see directly in an experiment (modulo the subsets of Lebesgue measure zero). 
Therefore the task is to reconstruct a wavefunction $\psi$ from a given pair of probability densities $(I, J)$. 
It turns out that the problem is quite complicated (W. Pauli himself had not given a complete answer). 
A generalization for a $d$-dimensional quantum system is straightforward. 
It is also of interest to point out that 
there exists a \emph{tomographic} generalization of this problem 
(for a review, see \cite{IbortMankoMarmocSimonicVentrigliac}): 
one is required to reconstruct a density operator $\wh{\rho}$ on $L^2 (\mathbb{R})$, 
$\wh{\rho}^{\dagger} = \wh{\rho} \geqslant 0$, 
$\mathrm{Tr} \wh{\rho} = 1$, rather than a wavefunction $\psi$, 
based on the marginal densities $I_{\mu, \nu} (z)$, $z \in \mathbb{R}$, corresponding to the observables 
$\mu Q + \nu P$,where  $\mu$ and $\nu$ vary over $\mathbb{R}$. 
This problem is known to have a solution. 

It turns out that a solution of the Pauli problem to construct $\psi$ out of a given 
pair $(I, J)$ does not need to be unique. Take as an example the following function: 
\begin{equation*} 
\psi_{a, b, c} (q) = (2 a/ \pi)^{1/ 4} \exp (- i (a + i b) q^2 + i c q), 
\end{equation*}
where $a \in \mathbb{R}_{> 0}$, $b \in \mathbb{R} \backslash \lbrace 0 \rbrace$, $c \in \mathbb{R}$. 
It is straightforward to check that 
\begin{equation*} 
|\psi_{a, b, c} (q)|^2 = |\psi_{a, - b, c} (q)|^2, \quad 
|\wt{\psi}_{a, b, c} (p)|^2 = |\wt{\psi}_{a, - b, c} (p)|^2,  
\end{equation*}
for $q, p \in \mathbb{R}$. 

The main idea of the present paper is to consider the experimental data about the fluctuations of 
internal energy $\delta U$ and inverse absolute temperature $\delta \beta$ as an \emph{input} data for the Pauli problem. 
More precisely, we may assume that we are given  
a pair of probability densities $w_{\delta U} (x)$ and $w_{\delta \beta} (y)$ of a 
more general shape than the Gauss exponents $f_{\delta U} (x)$ and $f_{\delta \beta} (y)$. 
The aim is to span a ``wavefunction'' over $w_{\delta U} (x)$ and $w_{\delta \beta} (y)$. 
In the main part of the paper I investigate when this is actually possible, i.e. when does a \emph{thermodynamic wavefunction} exist. 
It turns out that typically such a function can not be constructed, but on the other hand this only implies that the 
corresponding state is not pure. 
In other words, for a complete picture one needs to investigate a $d$-dimensional 
quantum tomographic problem in thermodynamics.

On the other hand, a thermodynamic wavefunction exists if we assume that the fluctuations are described 
by the Gauss exponents $f_{\delta U} (x)$ and $f_{\delta \beta} (y)$ \emph{precisely}. 
Furthermore, it is known that the quasithermodynamic fluctuation theory based on these exponents is quite useful in practice 
and really works. 
Therefore it is natural to consider an analogue of a semiclassical asymptotics $\hbar \to 0$ which in this case we may denote as $2 k_B \to 0$. 
In the second part of the paper I study the asymptotics of the thermodynamic Pauli problem 
corresponding to a vanishing Boltzmann constant 
and formulate a sufficient condition of existence of asymptotic solutions. 
The non-uniqueness of a solution in this case yields, in particular, a new mathematical 
generalization of the quasithermodynamic fluctuation theory. 

The general motivation for the present work comes from the idea to 
``\emph{quantize}'' thermodynamics 
in the sense of V.P.Maslov \cite{Maslov_thermo1, Maslov_thermo2, Maslov_thermo3, Maslov_ultra, Maslov_Nazaikinskii}. 
Note that in \cite{Maslov_op_meth} the ``thermodynamic'' terminology (enthalpy, entropy, etc.) is 
used to describe a (non-linear) complex germ on a Lagrangian manifold. 
One can look at the Lagrangian manifolds which emerge in thermodynamics and apply 
the semiclassical methods (the tunnel canonical operator \cite{Maslov_asymp_meth}) to study the 
asymptotics of the partition function of a system with respect to a large number of particles. 
A generalization of thermodynamics in this spirit is also of great interest in the context of quantum gravity 
\cite{Rovelli, ConnesRovelli, MontesinosRovelli, Rajeev_contact}.

Note that ``quantization'' of thermodynamics uses a different quantization parameter than $\hbar$. 
In this sense it is not exactly ``quantum'' thermodynamics (the thermodynamics of small systems)
where the aim is to construct a kind of mixture (a grand unification) of 
quantum mechanical and thermodynamic pictures 
\cite{AllahverdyanNieuwenhuizen, HenrichMichelMahler, QuanLiuSunNori, LindenPopescuShortWinter, SkrzypczykBrunnerLindenPopescu}
(i.e. both $\hbar$ and $k_B$ are involved). 
At the same time, a thermodynamic wavefunction 
in the sense as considered in the present paper (a ``deformation'' of quasithermodynamics)
is an extremely natural concept.

\section{Thermodynamic Lagrangian manifolds}
The concept of a Lagrangian manifold \cite{Maslov_op_meth, Maslov_asymp_meth} provides perhaps the most natural way to 
axiomatize \emph{phenomenological} thermodynamics. 
Note that in phenomenological thermodynamics we do \emph{not} have a notion of a ``number of particles'' 
and the Boltzmann constant $k_B$ is also not there. 
Instead, we work with the \emph{number of moles} $N$ of a chemical substance. 
In this sense, phenomenological thermodynamics is related to statistical thermodynamics 
like classical mechanics is related to quantum mechanics. 
The importance of this analogy is stressed in \cite{Maslov_thermo1}. 
The idea of ``quantization'' of energy (i.e. partitioning it into quanta) leads to an introduction of the Planck constant $\hbar$. 
Similarly, the idea of ``quantization'' of matter (i.e. partitioning it into corpuscles) leads to an introduction of the 
Boltzmann constant $k_B$. 
The number of particles (corpuscles) in one mole is the Avogadro number, 
\begin{equation*} 
N_{A} = R/ k_B, 
\end{equation*}
where $R$ is the universal gas constant -- a physical constant present at the level of \emph{phenomenological} thermodynamics. 

Take a simple one-component thermodynamic system described by absolute temperature $T$, volume $V$, and number of moles $N$. 
Denote: $U$ -- the internal energy, $S$ -- the entropy, $p$ -- the pressure, and $\mu$ -- the chemical potential. 
Then on the states of thermodynamic equilibrium we have: 
\begin{equation*} 
d U = T \,d S - p \,d V + \mu \,d N.
\end{equation*}
It is convenient to denote 
\begin{equation*} 
\beta := \frac{1}{T}, \quad 
\wt{p} := \frac{p}{T}, \quad 
\wt{\mu} := \frac{- \mu}{T}, 
\end{equation*}
and to rewrite this equation as 
\begin{equation*} 
d S = \beta \, d U + \wt{p} \, d V + \wt{\mu} \, d N. 
\end{equation*}
Assume that $(U, V, N)$ varies over a domain $D \subseteq \mathbb{R}^{3} (U, V, N)$ 
and that $S$ is described as a smooth function $S = S (U, V, N)$. 
Then we obtain a 3-dimensional surface $\Lambda \subset \mathbb{R}^{6} (\beta, \wt{p}, \wt{\mu}, U, V, N)$
described by the equations: 
\begin{equation*} 
\beta = \frac{\partial S (U, V, N)}{\partial U}, \quad 
\wt{p} = \frac{\partial S (U, V, N)}{\partial V}, \quad 
\wt{\mu} = \frac{\partial S (U, V, N)}{\partial N}.  
\end{equation*}
It is a common practice to write these derivatives as 
$(\partial S/ \partial U)_{V. N}$, $(\partial S/ \partial V)_{U. N}$, and 
$(\partial S/ \partial N)_{U. V}$, respectively. 
The surface $\Lambda$ is a Lagrangian manifold with respect to the symplectic structure 
\begin{equation*} 
\omega := 
d \, \beta \wedge d \, U +  
d \, \wt{p} \wedge d \, V + 
d \, \wt{\mu} \wedge d \, N, 
\end{equation*}
i.e. the pull-back $i_{\Lambda}^{*} (\omega) = 0$, where $i_{\Lambda}: \Lambda \to \mathbb{R}^{6} (\beta, \wt{p}, \wt{\mu}, U, V, N)$ is the 
canonical embedding. 
The first law of thermodynamics says basically that the differential 1-form $\alpha$ on $\Lambda$, 
\begin{equation*} 
\alpha := i_{\Lambda}^{*} (\beta \, d U + \wt{p} \, d V + \wt{\mu} \, d N), 
\end{equation*}
is \emph{exact}. 
The entropy $S$ is an \emph{action} on this Lagrangian manifold $\Lambda$ and it is measured in the same units as $k_B$: 
\begin{equation*}
[S] = [k_B]. 
\end{equation*}
Note that this is totally similar to mechanics: the mechanical action $A$ of a system is measured in the same units as $\hbar$, 
\begin{equation*} 
[A] = [\hbar], 
\end{equation*}
and if we write $A$ as a function of coordinates $q = (q_0, q_1, \dots, q_{n - 1}) \in \mathbb{R}^{n}$ 
(where $n$ is the number of degrees of freedom), $A = A (q)$, 
then we obtain a Lagrangian manifold 
$L \subset \mathbb{R}^{2 n} (p, q)$, $p = (p_0, p_1, \dots, p_{n - 1})$, 
described by the equations $p_i = \partial A (q)/ \partial q_i$, $i = 0, 1, \dots, n - 1$ 
(the symplectic structure is given by $\sum_{i = 0}^{n - 1} d p_i \wedge d q_i$). 

Let us define an \emph{abstract} thermodynamic system as follows. 
Consider a phase space $\mathbb{R}^{2 (d + 1)} (\beta, E)$, 
$\beta = (\beta_0, \beta_1, \dots, \beta_{d})$, $E = (E_0, E_1, \dots, E_{d})$, equipped with a symplectic 
structure 
\begin{equation*} 
\omega = \sum_{i = 0}^{d} d \beta_i \wedge d E_i. 
\end{equation*}
In the model example considered above $d = 2$, and we may put 
$\beta_0 = 1/ T$, $\beta_1 = p/ T$, $\beta_2 = - \mu/ T$, and 
$E_0 = U$, $E_1 = V$, $E_2 = N$. 
We identify an abstract thermodynamic system with a Lagrangian manifold (possibly, with a border): 
\begin{equation*} 
\Lambda \subset \mathbb{R}^{2 (d + 1)} (\beta, E), 
\end{equation*}
and denote by $i_{\Lambda}: \Lambda \to \mathbb{R}^{2 (d + 1)} (\beta, E)$ the canonical embedding 
of the system into the thermodynamic phase space $(\mathbb{R}^{2 (d + 1)} (\beta, E), \omega)$.

The conditions of $\Lambda$ are as follows: 
\begin{itemize} 
\item[(i)] 
The manifold $\Lambda$ is connected, simply connected, and admits $E = (E_0, E_1, \dots, E_d)$ as global coordinates. 

\item[(ii)] 
For any $\lambda > 0$ and any $a \in \Lambda$, there exists a point $b \in \Lambda$ such that 
$(\beta (b), E (b)) = (\beta (a), \lambda E (a))$. 
\end{itemize}
Here we write 
$(\beta (a), E (a)) \in \mathbb{R}^{2 (d + 1)} (\beta, E)$
for the coordinates of a point $a \in \Lambda$ acquired in the ambient space, 
and the notation $\lambda E$ is just 
 $(\lambda E_0, \lambda E_1, \dots, \lambda E_d)$, for $\lambda \in \mathbb{R}$, and 
$E = (E_0, E_1, \dots, E_d) \in \mathbb{R}^{d + 1} (E)$. 
The point $b \in \Lambda$ corresponding to a given 
$a \in \Lambda$ and $\lambda > 0$ in the condition (ii) is unique, and we denote it $b = \lambda a$. 
The coordinates $\beta \in \mathbb{R}^{d + 1} (\beta)$ are termed the \emph{intensive} coordinates of the system, 
and the coordinates $E \in \mathbb{R}^{d + 1} (E)$ are termed its \emph{extensive} coordinates.

\begin{prop} 
Assume the conditions (i) and (ii) are satisfied. 
Then there exists a \emph{canonical} choice of an action function $\wh{S}: \Lambda \to \mathbb{R}$, 
\begin{equation*}
d \wh{S} = i_{\Lambda}^{*} \Big( \sum_{i = 0}^{d} \beta_i \, d E_{i} \Big), 
\end{equation*}
determined by the condition 
\begin{equation*} 
\wh{S} (\lambda a) = \lambda \wh{S} (a), 
\end{equation*}
for any $\lambda > 0$ and $a \in \Lambda$. 
\end{prop}

\emph{Proof.}
Fix an arbitrary point $a_0 \in \Lambda$. 
Any solution of the equation for $\wh{S}$ is of the form: 
$\wh{S} (a) := \varkappa + \int_{\gamma} \sum_{i = 0}^{d} \beta_i \, d E_i$, 
where $\gamma \subset \Lambda$ is a path from $a_0$ to $a$, and $\varkappa \in \mathbb{R}$. 
Such a path always exists, since $\Lambda$ is connected. 
The value of the integral does not change under continuous deformations of $\gamma$ 
since $\Lambda$ is a Lagrangian manifold, 
and, furthermore, it does not depend on a choice of $\gamma$ since $\Lambda$ is simply connected. 
Therefore the space of solutions of the equation for $\wh{S}$ is parametrized by $\varkappa \in \mathbb{R}$. 

Since $\Lambda$ admits global coordinates $E = (E_0, E_1, \dots, E_d)$, we can define a function 
$S = S (E)$ such that 
$\wh{S} (a) = S (E (a))$, for any $a \in \Lambda$. 
The condition (ii) implies that $\beta_i (a) = \beta_i (\lambda a)$, $i = 0, 1, \dots, d$, 
for any $a \in \Lambda$, and $\lambda > 0$. 
Since $\beta_i (a) = (\partial S (E)/ \partial E_i)|_{E = E (a)}$, $i = 0, 1, \dots, d$, 
we obtain a system of equations 
\begin{equation*} 
\frac{\partial}{\partial E_i} (\lambda^{-1} S (\lambda E) - S (E)) = 0, 
\end{equation*}
where $i = 0, 1, \dots, d$. It follows that 
\begin{equation*} 
S (\lambda E) = \lambda S (E) + c (\lambda), 
\end{equation*}
where $c (\lambda)$ is a differentiable function, $\lambda > 0$. 
If we take another real parameter $\mu > 0$, then, applying the scaling twice, we obtain: 
on one hand, $S (\lambda \mu E) = \lambda \mu S (E) + c (\lambda \mu)$, and, on the other hand, 
$S (\lambda \mu E) = \lambda [\mu S (E) + c (\mu)] + c (\lambda)$. 
Therefore, we have an equation: 
\begin{equation} 
c (\lambda \mu) = \lambda c (\mu) + c (\lambda), 
\label{eq:constant_entropy}
\end{equation}
where $\lambda, \mu > 0$. 
Differentiating it with respect to $\mu$ and cancelling out the factor $\lambda > 0$ yields: $c' (\lambda \mu) = c' (\mu)$, 
i.e. $c'$ is a constant function. Then $c (\mu) = p \mu + q$, where $p, q \in \mathbb{R}$, and 
substituting it into the equation 
\eqref{eq:constant_entropy}
yields: 
$p \lambda \mu + q = \lambda [p \mu + q] + (p \lambda + q)$. It follows that $\lambda (p + q) = 0$, i.e. $p = -q$, and we 
have: $S (\lambda E) = \lambda S (E) + q (- \lambda + 1)$. 
Hence: 
\begin{equation*} 
S (\lambda E) - q = \lambda (S (E) - q).  
\end{equation*}
The function $\wh{S}_{0} (a) := S (E (a)) - q$ satisfies 
$\wh{S}_0 (\lambda a) = \lambda \wh{S}_0 (a)$, where $a \in \Lambda$, $\lambda > 0$, 
and it is a unique function with such a property, since all other variants $\wh{S} (a)$ differ from $\wh{S}_{0} (a)$ by a 
constant. \qed

\begin{dfn} 
The real function $\wh{S}: \Lambda \to \mathbb{R}$ 
satisfying the conditions of the proposition 1 
is termed an \emph{entropy} 
of an abstract thermodynamic system $\Lambda$. 
\end{dfn}

\vspace{0.2 true cm}
\noindent
\emph{Remark.}
We do not have a 
``shift of a reference point'' relative to which we perceive the entropy of an abstract thermodynamic system. 
The entropy is canonically defined 
as long as we accept the \emph{extensivity} of our theory (the condition (ii)). 
\hfill $\Diamond$
\vspace{0.2 true cm}

It is now the right moment to formulate the third law of thermodynamics, 
which is basically a \emph{condition on $\Lambda$ at infinity}: 
\begin{itemize} 
\item[(iii)] 
For any sequence of points $\lbrace a_{n} \rbrace_{n = 0}^{\infty} \subset \Lambda$, such that 
the inverse absolute temperature $\beta_0 (a_n) \to + \infty$, 
while $E_i (a_{n}) = E_i (a_0)$, $i = 1, 2, \dots, d$, as $n \to \infty$, 
there exists a finite limit $\lim_{n \to \infty} \wh{S} (a_n)$, and 
its value is the same for every such sequence. 
\end{itemize}

\vspace{0.2 true cm}
\noindent
\emph{Remark.} 
The fact that the limit in (iii) is actually the same has quite serious implications 
(the Nernst theorem). For any $a_0 \in \Lambda$, 
if $\gamma = \lbrace a (t) \rbrace_{t \in [0, + \infty[} \subset \Lambda$ is a continuous curve (a process) 
starting at $a (0) = a_0$ and  
approaching the absolute zero of temperature, $\beta_0 (a (t))^{-1} \to 0$, as $t \to + \infty$, 
then it cannot be realized as a \emph{finite} union $\gamma = \cup_{i = 0}^{n - 1} \gamma_{i}$, $n \in \mathbb{Z}_{> 0}$, 
of isothermic and adiabatic parts 
(recall that $\gamma_i$ is termed \emph{isothermic} iff
$\beta_0 (a)^{-1}$ is constant along $a \in \gamma_i$, 
and $\gamma_i$ is termed \emph{adiabatic} iff 
$\wh{S} (a)$ is constant along $a \in \gamma_i$). 
Intuitively, and adiabatic process is a ``swift'' change of the values of extensive quantities, 
and the isothermic process corresponds to another extreme: it is very ``slow''. 
An informal statement of the Nernst theorem has a deep flavour of ancient Greek philosophy: 
it is impossible to reach the absolute zero in a \emph{finite} number of steps.  
\hfill $\Diamond$
\vspace{0.2 true cm}

Let $\wh{S} = \wh{S} (a)$ be the entropy on $\Lambda$, and let $S = S (E)$ be a function 
determined by $S (E (a)) = \wh{S} (a)$, for every $a \in \Lambda$, (entropy as a function of extensive coordinates). 
Let $I \subset \lbrace 0, 1, \dots, d \rbrace$. Denote 
$\bar I := \lbrace 0, 1, \dots, d \rbrace \backslash I$, and write: 
\begin{equation*}
I = \lbrace i_0 < i_1 < \dots < i_{n - 1} \rbrace, \quad 
\bar I = \lbrace i_{n} < i_{n + 1} < \dots < i_{d} \rbrace, 
\end{equation*}
where $n = |I|$ is the cardinality of $I$. 
For a phase space point $(\beta, E)$, put 
$\beta_{I} := (\beta_{i_0}, \beta_{i_1}, \dots, \beta_{i_{n - 1}})$ and 
$E_{\bar I} := (E_{i_n}, E_{i_{n + 1}}, \dots, E_{i_d})$, 
and define a projection 
\begin{equation*} 
\wh{\pi}_{I}: \mathbb{R}^{2 (d + 1)} \to \mathbb{R}^{d + 1}, \quad 
(\beta, E) \mapsto (\beta_{I}, E_{\bar I}). 
\end{equation*}
Note that due to the condition (i) we have: $\dim (\wh{\pi}_{\emptyset} \Lambda) = d + 1$, but the 
condition (ii) implies: $\dim (\wh{\pi}_{\bar \emptyset} \Lambda) \leqslant d$. 

Take a point $a_0 \in \Lambda$ and a subset $I \subset \lbrace 0, 1, \dots, d \rbrace$. 
Assume that there exists a neighbourhood $U \subset \Lambda$ of the point $a_0 \in U$, such that 
$\dim (\wh{\pi}_{I} U) = d + 1$. 
Assume that if $a \in U$ then $\lambda a \in U$, for every $\lambda > 0$. 
Then we can consider $(\beta_{I}, E_{\bar I})$ as local coordinates around $a_0 \in \Lambda$ 
and define a function $S_{U}^{I}: \wh{\pi}_{I} (U) \to \mathbb{R}$ as follows: 
\begin{equation*} 
S_{U}^{I} (\beta_I (a), E_{\bar I} (a)) = \wh{S} (a) - \sum_{i \in I} \beta_{i} (a) E_{i} (a), 
\end{equation*} 
for every $a \in U$. 
Observe that $S_{U}^{I} (\beta_{I} (a), \lambda E_{\bar I} (a)) = S_{U}^{I} (\beta_{I} (\lambda a), E_{\bar I} (\lambda a))$, 
where $\lambda > 0$, and therefore, the extensivity $\wh{S} (\lambda a) = \lambda \wh{S} (a)$ implies:  
\begin{equation*} 
S_{U}^{I} (\beta_{I}, \lambda E_{\bar I}) = \lambda S_{U}^{I} (\beta_{I}, E_{\bar I}), 
\end{equation*}
for any $(\beta_{I}, E_{\bar I}) \in \wh{\pi}_{I} (U)$ and $\lambda > 0$. 
The neighbourhood $U \subset \Lambda$ is described by the equations: 
\begin{equation*} 
E_{i} = - \partial S_{U}^{I} (\beta_{I}, E_{\bar I})/ \partial \beta_i, \quad 
\beta_{j} = \partial S_{U}^{I} (\beta_{I}, E_{\bar I})/ \partial E_j, 
\end{equation*}
where $i \in I$, $j \in \bar I$, and $(\beta_{I}, E_{\bar I}) \in \wh{\pi}_{I} (U)$. 

Consider a particular case $I = [d]$, where 
\begin{equation*} 
[d] := \lbrace 0, 1, \dots, d - 1 \rbrace, 
\end{equation*}
and assume that $E_{d} (a) > 0$ for all $a \in U$ (for example, $E_d$ is the volume of the system $V$). 
Then we obtain $S_{U}^{[d]} (\beta_{[d]}, E_{d}) = E_{d} S_{U}^{[d]} (\beta_{[d]}, 1)$, and 
\begin{equation*} 
\beta_{d} = S_{U}^{[d]} (\beta_{[d]}, 1). 
\end{equation*}
The last equation can be perceived as a description of the projection 
$\wh{\pi}_{\bar \emptyset} (U) \subset \mathbb{R}^{d + 1} (\beta)$, 
$\dim \wh{\pi}_{\bar \emptyset} (U) = d$. 
The way to recover $U$ from the projection $\wh{\pi}_{\bar \emptyset} (U)$ is given by the formulae: 
$E_i/ E_{d} = - \partial S_{U}^{[d]} (\beta_{[d]}, 1)/ \partial \beta_{i}$, $i \in [d]$.

\section{Reduction of degrees of freedom}

If we have a collection of abstract thermodynamic systems 
$\Lambda_{\alpha} \subset \mathbb{R}^{2 (d_{\alpha} + 1)} (\beta^{(\alpha)}, E^{(\alpha)})$, $d_{\alpha} \in \mathbb{Z}_{> 0}$, 
$\alpha = 0, 1, \dots, n - 1$, then it is possible to construct other thermodynamic systems out of it. 
The most simple case is where the systems do not interact with each other. 
Then together they form a system (the \emph{direct product}) in a phase space $\mathbb{R}^{2 (d + 1)} (\beta, E)$, 
$d = d_0 + d_1 + \dots + d_{n - 1}$,  with 
intensive coordinates  
$\beta := (\beta^{(0)}, \beta^{(1)}, \dots, \beta^{(n - 1)})$, and 
extensive coordinates 
$E := (E^{(0)}, E^{(1)}, \dots, E^{(n - 1)})$, 
described by a Lagrangian manifold 
\begin{equation*} 
\Lambda := \Lambda^{(0)} \times \Lambda^{(1)} \times \dots \times \Lambda^{(n - 1)}. 
\end{equation*}
The symplectic form is $\omega := \sum_{\alpha = 0}^{n - 1} \sum_{i = 0}^{d_{\alpha}} d \beta_{i}^{(\alpha)} \wedge d E_{i}^{(\alpha)}$, 
and the entropy $\wh{S} = \wh{S} (a)$ on $\Lambda$ is just a sum of the entropy functions: 
\begin{equation*} 
\wh{S} (a) = \sum_{\alpha = 0}^{n - 1} \wh{S}_{\alpha} (a_{\alpha}), 
\end{equation*}
where $a = (a_0, a_1, \dots, a_{n - 1}) \in \Lambda$, and $\wh{S}_{\alpha}$ is the entropy function on $\Lambda_{\alpha}$, 
$\alpha = 0, 1, \dots, n - 1$. 

If the subsystems $\Lambda_{\alpha}$, $\alpha = 0, 1, \dots, n - 1$, begin to interact, then 
we can perceive it as if some of the extensive degrees of freedom are being ``released''.  
Consider a linear non-degenerate transformation of extensive coordinates: 
\begin{equation*} 
E_{i}' = \sum_{j = 0}^{d} C_{i, j} E_{j}, 
\end{equation*}
where $i = 0, 1, \dots, d$, and $C = \| C_{i, j} \|_{i, j = 0}^{d} \in \mathit{GL}_{d+1} (\mathbb{R})$. 
This lifts to a canonical transformation 
$\omega = \sum_{i = 0}^{d} d \beta_i \wedge d E_i = \sum_{i = 0}^{d} d \beta_i' \wedge d E_i'$, 
if we put 
\begin{equation*} 
\beta_{i}' := \sum_{j = 0}^{d} (C^{-1})_{i, j} \beta_{j}, 
\end{equation*}
where $i = 0, 1, \dots, d$. 
As long as the subsystems do not interact, we can hold the values of the extensive coordinates $E_{0}', E_{1}', \dots, E_{d}'$. 
An interaction with respect to the last $d - d'$ coordinates, $0 \leqslant d' < d$, means that 
$E_{d' + 1}', E_{d' + 2}', \dots, E_{d}'$ are allowed to vary, so that the total system $\Lambda$ ends up 
in a state with a maximum entropy (the \emph{second law of thermodynamics}). 

Write the entropy function $\wh{S}$, $a \in \Lambda$, in terms of extensive coordinates $E'$: 
$\wh{S} (a) = S' (E' (a))$, $a \in \Lambda$. 
If we start in a point $a \in \Lambda$, 
and release the coordinates $E_{d' + 1}', E_{d' + 2}', \dots, E_{d}'$, 
then the final state $b = b (a)$ after a relaxation to an equilibrium satisfies: 
\begin{equation*}
E_i' (b (a)) = E_i' (a), \quad 
\Big( \frac{\partial S' (E')}{\partial E_{j}'} \Big) \Big|_{E' = E' (b (a))} = 0, 
\end{equation*}
where $i = 0, 1, \dots, d'$, and 
$j = d' + 1, d' + 2, \dots, d$. 

Since $\beta_i' (a) = (\partial S' (E')/ \partial E_i')|_{E' = E' (a)}$, $i = 0, 1, \dots, d$, 
we observe that $\beta_j' (b (a)) = 0$, $j = d' + 1, d' + 2, \dots, d$, and that 
the corresponding values of $E_{j}' (b (a))$, $j = d' + 1, d' + 2, \dots, d$, are determined by 
$E_{i}' (b (a))$, $i = 0, 1, \dots, d'$. 
This allows to define a ``reduced'' entropy $\bar S' = \bar S' (\bar E')$, $\bar E' = (E_{0}', E_{1}', \dots, E_{d'}')$ 
as follows: 
\begin{equation*} 
\bar S' (\bar E' (b (a))) = S' (E' (b (a))), 
\end{equation*}
for any $a \in \Lambda$. 
The values intensive coordinates $\beta_i' (b (a))$, $i = 0, 1, \dots, d'$, 
can be expressed not in terms of function $S' = S' (E')$, but it terms of the reduced entropy $\bar S' = \bar S' (\bar E')$: 
$\beta_{j}' (b (a)) = (\partial \bar S' (\bar E')/ \partial E_j')|_{E' = E' (b (a))}$. 
It follows that we obtain a Lagrangian manifold 
$\bar \Lambda' := \mathbb{R}^{2 (d' + 1)} (\bar \beta', \bar E')$, 
$\bar \beta' := (\beta_0', \beta_1', \dots, \beta_{d'}')$, 
described by the equations: 
\begin{equation*} 
\beta_i' = \partial \bar S' (\bar E')/ \partial E_{i}', 
\end{equation*}
where $i = 0, 1, \dots, d'$, and the symplectic structure is given by 
$\bar \omega' := \sum_{i = 0}^{d'} d \beta_{i}' \wedge d E_i'$. 
This manifold is nothing else but an image of 
\begin{equation*} 
M := \lbrace a \in \Lambda \,|\, \beta_{i} (a) = 0, \, i = 0, 1, \dots, d' \rbrace
\end{equation*}
under the canonical projection 
$\wh{\pi}: \mathbb{R}^{2 (d + 1)} (\beta', E') \to \mathbb{R}^{2 (d' + 1)} (\bar \beta', \bar E')$. 

In terms of the original coordinates $(\beta, E)$, 
the submanifold $M \subset \Lambda$ can be perceived as follows. 
We have a family of planes 
\begin{equation*} 
L_{p}^{(k)} := \lbrace E \in \mathbb{R}^{d + 1} \,|\, 
q_{0}^{(k)} E_{0} + q_{1}^{(k)} E_{1} + \dots + q_{d}^{(k)} E_{d} = p
\rbrace, 
\end{equation*}
where $p \in \mathbb{R}$, and the 
coefficients $q^{(k)} := (q_{0}^{(k)}, q_{1}^{(k)}, \dots, q_{d}^{(k)})$ form a $(d + 1)$-dimensional real unit vector, 
$k = 0, 1, \dots, d'$. 
We assume that the intersections 
\begin{equation*} 
L_{p_{0}, p_{1}, \dots, p_{d'}} := L_{p_{0}}^{(0)} \cap L_{p_{1}}^{(1)} \cap \dots \cap L_{p_{d'}}^{(d')}
\end{equation*}
are $(d - d')$-dimensional, where $(p_0, p_1, \dots, p_{d'}) \in \mathbb{R}^{d' + 1}$. 
Put 
\begin{equation*} 
\wt{L}_{p_0, p_1, \dots, p_{d'}} := \lbrace a \in \Lambda \,|\, E (a) \in L_{p_0, p_1, \dots, p_{d'}} \rbrace. 
\end{equation*} 
For every $a \in \Lambda$, we can compute $E (a)$, and then 
take the unique $L_{p_0 (a), p_1 (a), \dots, p_{d'} (a)}$ containing $E (a)$. 

The planes $L_{p}^{(k)}$, $p \in \mathbb{R}$, $k = 0, 1, \dots, d'$, 
define basically how the subsystems of $\Lambda$ interact with each other. 
Once we ``switch on'' the interaction, 
the initial point $a \in \Lambda$ starts to move remaining in the submanifold 
$\wt{L}_{p_0 (a), p_1 (a), \dots, p_{d'} (a)}$, until it finds a point $b = b (a)$ of maximal entropy. 
The collection of points $b (a)$, where $a$ varies over $\Lambda$ 
(the image of the retraction $a \mapsto b (a)$), yields $M$. 

\vspace{0.2 true cm}
\noindent
\emph{Example 1.} 
Consider two systems 
$\Lambda^{(0)}$ and $\Lambda^{(1)}$. 
with extensive coordinates $(U^{(\alpha)}, V^{(\alpha)}, N^{(\alpha)})$, $\alpha = 0, 1$,
(internal energy, volume, number of moles). 
If the two systems are put in a thermal contact, 
then their internal energies can change as follows: 
$E^{(0)} \to E^{(0)} + x$, $E^{(1)} \to E^{(1)} - x$, where $x \in \mathbb{R}$ is a parameter. 
The other extensive parameters (volumes and numbers of moles) in this case remain fixed. 

Assume that there is another channel for a change of state: 
$E^{(0)} \to E^{(0)} + y$, $E^{(1)} \to E^{(1)} - y$, $N^{1} \to N^{(1)} - q y$, where 
$y \in \mathbb{R}$ is a parameter, and $q > 0$ is a constant. 
Then the total picture is as follows: 
$E^{(0)} \to E^{(0)} + x + y$, 
$E^{(1)} \to E^{(1)} - (x + y)$, and 
$N^{(1)} \to N^{(1)} - q y$, 
while the other extensive quantities $N^{(0)}$, $V^{(0)}$, and $V^{(1)}$, are fixed. 
Observe that $E^{(0)} + E^{(1)}$ is also fixed. 
Define a linear transformation of extensive coordinates: 
\begin{equation*} 
(E_0', E_1', \dots, E_5') := (E^{(0)} + E^{(1)}, N^{(0)}, V^{(0)}, V^{(1)}, (E^{(0)} - E^{(1)})/ 2, N^{(1)}). 
\end{equation*}
The corresponding transformation of 
$(\beta^{(\alpha)}, \wt{p}^{(\alpha)}, \wt{\mu}^{(\alpha)})$, $\alpha = 0, 1$ 
(inverse absolute temperature, pressure over temperature, and minus chemical potential over temperature)
is of the shape: 
\begin{equation*} 
(\beta_0', \beta_1', \dots, \beta_5') := 
((\beta^{(0)} + \beta^{(1)})/ 2, \wt{\mu}^{(0)}, \wt{p}^{(0)}, \wt{p}^{(1)}, \beta^{(0)} - \beta^{(1)}, \wt{\mu}^{(1)}). 
\end{equation*}
Reduce the thermodynamic system $\Lambda = \Lambda^{(0)} \times \Lambda^{(1)}$ with respect to the last two degrees of freedom in the transformed coordinates. 
A point $a \in \Lambda$ retracts to a point $b = b (a)$ satisfying 
the conditions $\beta_4' (b (a)) = 0$, and $\beta_5' (b (a)) = 0$. 
We conclude: 
\begin{equation*} 
\beta^{(0)} - \beta^{(1)} = 0, \quad 
\wt{\mu}^{(1)} = 0, 
\end{equation*}
in a state of thermodynamic equilibrium. 
This example mimics the black body radiation in a thermostat: 
the temperature of the radiation acquires the temperature of the thermostat, and 
its chemical potential vanishes. 
\hfill $\Diamond$
\vspace{0.2 true cm}

\vspace{0.2 true cm}
\noindent
\emph{Example 2.} 
Consider a system $\Lambda$ consisting of three chemical substances ($A$, $B$, and $C$) with an entropy function 
$S = S (U, V, N_0, N_1, N_2)$, where $U$ is the internal energy, $V$ is volume, and $N_{0}$, $N_{1}$, $N_{2}$ 
are the number of moles of the substances $A$, $B$, and $C$, respectively. 
Consider the following chemical reaction: 
\begin{equation*} 
A + B \rightleftarrows 2 C. 
\end{equation*}
Note that the stoichiometric coefficients are well-defined already on the level of \emph{phenomenological} thermodynamics. 
The number of moles can undergo the following changes: $N_0 \to N_0 - x$, $N_1 \to N_1 - x$, $N_2 \to N_2 + 2 x$, 
where $x \in \mathbb{R}$ is a parameter. 
We observe that the quantities $D := N_0 - N_1$ and $N := N_0 + N_1 + N_2$ are fixed. 
Introduce the extensive coordinates $(E_0', E_1', \dots, E_4') = (U, V, D, N, N_3)$ and reduce the system with respect to $N_3$. 
We have: $N_0 = (N + D - N_3)/2$, and $N_1 = (N - D - N_3)/2$, 
so the intensive quantity $\beta_{4}'$ corresponding to $E_4' = N_3$ is of the shape: 
$\beta_4' = - (\wt{\mu}_1 + \wt{\mu}_2)/2 + \wt{\mu}_3$, 
where $\wt{\mu}_0$, $\wt{\mu}_1$, and $\wt{\mu}_2$, are the intensive coordinates corresponding to $N_1$, $N_2$, $N_3$, respectively 
(the minus chemical potentials over the absolute temperature). 
If we start in a point $a \in \Lambda$, then 
the final point $b = b (a)$ satisfies $\beta_4' (b (a)) = 0$, i.e. we obtain an equation 
\begin{equation*} 
- \wt{\mu}_{0} - \wt{\mu}_1 + 2 \wt{\mu}_2 = 0, 
\end{equation*}
describing a 4-dimensional submanifold $M \subset \Lambda$. 
In physical chemistry one obtains different characterizations of this surface depending an a chosen model. 
Consider, for instance, the following one: 
$U (b (a)) = U (a)$, $V (b (a)) = V (a)$, and 
the values of $N_0 (b (a))$, $N_1 (b (a))$, and $N_2 (b (a))$, are of the shape 
$N_0 (b (a)) = N_0 (a) - x$, $N_1 (b (a)) = N_1 (a) - x$, and 
$N_2 (b (a)) = N_2 (a) + 2 x$, 
with $x$ being a solution of the equation 
\begin{equation} 
K = \frac{(N_2 (a) + 2 x)^2}{(N_0 (a) - x) (N_1 (a) - x)}, 
\label{eq:chem_equilib}
\end{equation}
where $K \in \mathbb{R}_{> 0}$ is the \emph{chemical equilibrium constant} 
(the powers of the factors in the numerator and the denominator are the stoichiometric coefficients). 
In our case this is just a quadratic equation on $x$. 
If $K \to +0$, then in the limit we obtain: 
$(2 x + N_2 (a))^2 = 0$, i.e. $x = - N_2 (a)/2$, 
so the limit of the entropy (as a function $\wh{S}$ on the manifold $\Lambda$) in the final state $b (a)$ is of the shape: 
\begin{equation*} 
\wh{S} (b (a)) = S (U (a), V (a), N_0 (a) + N_2 (a)/ 2, N_1 (a) + N_2 (a)/2, 0), 
\end{equation*}
i.e. the chemical equilibrium is shifted completely to the left. 
On the other hand, if $K \to + \infty$, then in the limit we have: 
$(x - N_0 (a)) (x - N_1 (a)) = 0$. Since the quantities $N_0 (a) - x$ and $N_1 (a) - x$ must be non-negative, 
we obtain: $x = \min \lbrace N_0 (a), N_1 (a) \rbrace$. 
It follows that $N_0 (b (a)) = 0$, if $N_0 (a) < N_1 (a)$, and that 
$N_1 (b (a)) = 0$, if $N_0 (a) > N_1 (a)$. 
The chemical equilibrium is shifted completely to the right. 
If $N_0 (a) = N_1 (a)$ then 
\begin{equation*} 
\wh{S} (b (a)) = S (U (a), V (a), 0, 0, N_0 (a) + N_1 (a) + N_2 (a)), 
\end{equation*}
i.e. $A$ and $B$ have reacted completely 
(since there was no excess of any of the substances) and 
everything has turned into $C$. 
\hfill $\Diamond$
\vspace{0.2 true cm}

\vspace{0.2 true cm}
\noindent
\emph{Example 3.} 
Consider a system $\Lambda = \Lambda^{(0)} \times \Lambda^{(1)}$ which is a direct product of two copies 
of the system considered in the previous example. 
We have the extensive coordinates $(U^{(\alpha)}, V^{(\alpha)}, N_{0}^{(\alpha)}, N_{1}^{(\alpha)}, N_{2}^{(\alpha)})$ for 
each sybsystem $\alpha = 0, 1$. 
Assume first that the chemical reaction is totally suppressed and that the subsystems can exchange 
the internal energies and the chemical substances. 
Introduce the extensive coordinates 
$U := U^{(0)} + U^{(1)}$, and $N_{i} := N_{i}^{(0)} + N_{i}^{(1)}$, $i = 0, 1, 2$. 
The reduction with respect to the coordinates $D := (U^{(0)} - U^{(1)})/ 2$ and 
$D_i := (N_{i}^{(0)} - N_{i}^{(1)})/ 2$, $i = 0, 1, 2$, yields a submanifold $M \subset \Lambda$ 
described by the equations: 
$\beta_{0}^{(0)} - \beta_{0}^{(1)} = 0$, and 
$\wt{\mu}_{i}^{(0)} - \wt{\mu}_{i}^{(1)} = 0$, $i = 0, 1, 2$, 
i.e. the temperatures and the chemical potentials of the substances become equal in the subsystems. 
If we take $a \in \Lambda$, then after a relaxation to equilibrium we obtain $b (a) \in M$. 
The restriction of the entropy function $\wh{S}|_{M}$ written in terms of the coordinates 
$(U, V^{(0)}, V^{(1)}, N_0, N_1, N_2)$ yields a generating function $\bar S = \bar S (U, V^{(0)}, V^{(1)}, N_0, N_1, N_2)$ 
of the Lagrangian manifold $\bar \Lambda$ of the reduced system.

Switch on the chemical reaction $A + B \rightleftarrows 2 C$ (the chemical equilibrium constant is $K$). 
In accordance with formula \eqref{eq:chem_equilib} of the previous example, our point $b (a) \in M$ should 
retract now to a point $c (a) \in \bar M$ 
on a submanifold $\bar M \subset \bar \Lambda$ described by the equation 
$N_2^2/ (N_0 N_1) = K$. 

It is quite of interest to consider the following special case (the \emph{Gibbs paradox}). 
Let $a \in \Lambda$ correspond to 
\begin{equation*}
\begin{gathered}
(N_{0}^{(0)} (a), N_{1}^{(0)} (a), N_{2}^{(0)} (a)) = (n, 0, 0), \\
(N_{0}^{(1)} (a), N_{1}^{(1)} (a), N_{2}^{(1)} (a)) = (0, n, 0), 
\end{gathered}
\end{equation*}
where $n \in \mathbb{R}_{> 0}$. 
Let $V^{(0)} (a) = V^{(1)} (a) = v$ and $U^{(0)} (a) = U^{(1)} (a) = u$, where $u, v \in \mathbb{R}_{> 0}$. 
Denote the entropy corresponding to the initial point $a \in \Lambda$ as $S_{\mathit{in}} (a)$. 
Denote the entropy corresponding to the final point $c (a) \in \bar M$ as $S_{\mathit{out}} (a; K)$. 
We have: 
\begin{equation*} 
S_{\mathit{in}} (a) = S (u, v, n, 0, 0) + S (u, v, 0, n, 0), 
\end{equation*}
where $S$ is the entropy as a function of extensive coordinates from the previous example. 
At the same time: 
\begin{equation*} 
\lim_{K \to + 0} S_{\mathit{out}} (a; K) = S (2 u, 2 v, n, n, 0), 
\end{equation*}
and 
\begin{equation*} 
\lim_{K \to + \infty} S_{\mathit{out}} (a; K) = S (2 u, 2 v, 0, 0, 2 n). 
\end{equation*}
The mixing entropy 
\begin{equation*} 
S_{\mathit{mix}} (a; K) := S_{\mathit{out}} (a; K) - S_{\mathit{in}} (a)
\end{equation*}
can vary depending on the parameter $K$. 
Suppose now than the chemical substances $A$ and $B$ are very-very similar. 
For example, we measure this similarity in terms of molar masses $M_0$ and $M_1$, respectively. 
Let the chemical reaction $A + B \rightleftarrows 2 C$ consist in creating a substance $C$ 
with a molar mass $M_2 = (M_0 + M_1)/ 2$. Put 
\begin{equation*} 
\varepsilon := |M_0 - M_1|/ M_2. 
\end{equation*}
Let $K = K (\varepsilon)$ smoothly depend on this parameter in such a way that 
$K (\varepsilon) \ll 1$, if $\varepsilon < \varepsilon_0/ 2$, and 
$K (\varepsilon) \gg 1$, if $\varepsilon > \varepsilon_0$, where $\varepsilon_0 \in \mathbb{R}_{> 0}$. 

Fix $M_2 = M \in \mathbb{R}_{> 0}$, and denote the entropy function of the chemical substance with this molar mass 
as $S_M = S_M (U, V, N)$, where $U$ is the internal energy, $V$ is the volume, and $N$ is the number of moles. 
If $\varepsilon < \varepsilon_0/ 2$, then we do not ``see'' a difference between the substances $A$, $B$, and $C$, and 
therefore: 
\begin{equation*} 
S_{\mathit{mix}} (a; K (\varepsilon)) \approx S_{M} (2 u , 2 v, 2 n) - 2 S_{M} (u, v, n) = 0, 
\end{equation*}
due to the \emph{extensivity} of the entropy function. 
On the other hand, if $\varepsilon > \varepsilon_0$, then we can clearly state that $A$, $B$, and $C$, are different, 
and there appears an observable jump $S_{\mathit{mix}} (a; K (\varepsilon)) > 0$. 
If we put $S (2 u, 2 v, n, n, 0) = 2 S_{M} (u, 2 v, n)$, and take 
\begin{equation*} 
S_M (U, V, N) = k_B N \Big\lbrace 
\log \Big[ 
\Big( \frac{k_B}{R} \Big)^{7/ 2} \frac{4 \pi M}{3 (2 \pi \hbar)^{2}} \frac{V}{N} 
\Big( \frac{U}{N} \Big)^{3/ 2} \Big] + \frac{5}{2} \Big\rbrace, 
\end{equation*}
(the Suckur-Tetrode equation for $N$ moles of a monoatomic gas in three dimensions), 
then in the region $\varepsilon > \varepsilon_0$ we obtain: 
\begin{equation*} 
S_{\mathit{mix}} (a; K (\varepsilon)) \approx 2 k_{B} N \log (2).
\end{equation*}
This is precisely the jump of entropy considered in many discussions about the Gibbs paradox. 
The present example mimics a resolution of the Gibbs paradox in statistical physics suggested in \cite{Maslov_Gibbs}. 
This resolution is based on a construction of a kind of ``number-theoretic'' Bose gas of 
fractional dimension $d = 2 \gamma + 1$, $0 < \gamma \leqslant 1$, but loosely speaking the philosophy can be reformulated as follows. 
Take a gas of identical ``red'' particles and a similar gas of identical ``blue'' particles 
(assume that all other parameters like mass, size, etc. of the particles are the same). 
Then the result of mixing is not a gas of particles some of which are ``red'', and some of which are ``blue''. 
The ``correct'' answer: the result is a gas of \emph{identical} ``purple'' particles. 
These colours correspond to different dimensions of the number-theoretic Bose gas. 
In the present example, 
the substance $A$ is, for instance, ``red'', the substance $B$ is ``blue'', and the substance $C$ is ``purple''. 
The chemical reaction $A + B \rightleftarrows 2 C$ can be perceived as a \emph{``loss of identity''}. 
\hfill $\Diamond$
\vspace{0.2 true cm}

\section{Quasithermodynamic fluctuations}
Take an abstract thermodynamic system $\Lambda \subset \mathbb{R}^{2 (d + 1)} (\beta, E)$ 
(a Lagrangian manifold with respect to the canonical symplectic structure 
$\sum_{i = 0}^{d} d \beta_{i} \wedge d E_{i}$ satisfying the conditions (i), (ii), and (iii)). 
A reduction with respect to the extensive coordinates $E_{d' + 1}, E_{d' + 2}, \dots, E_{d}$, 
where $d' < d$ defines a retraction $\Lambda \to M$, $a \mapsto b (a)$, to a 
submanifold $M  \subset \Lambda$ of dimension $\dim M = d' + 1$. 

In \emph{quasithermodynamics} the reduced thermodynamic quantities $E_{j}$, $j = d' + 1, d' + 2, \dots, d$, 
are not exactly fixed, but they \emph{fluctuate} around the equilibrium values 
$E_j (b)$, $j = d' + 1, d' + 2, \dots, d$, where $a \in M$. 
The fluctuations are described by a collection of random variables 
$\delta E_{[d' + 1, d]} := (\delta E_{d' + 1}, \delta E_{d' + 2}, \dots, \delta E_{d})$. 
It is convenient to introduce the following concept. 
\begin{dfn} 
An abstract thermodynamic system $\Lambda \subset \mathbb{R}^{2 (d + 1)} (\beta, E)$ 
with an entropy function $S = S (E)$ is termed \emph{linearly stable} iff 
for every $I \subset \lbrace 0, 1, \dots, d \rbrace$, 
$1 \leqslant |I| \leqslant d$, and $a \in \Lambda$, 
the symmetric matrix obtained from 
the symmetric matrix 
$\| (\partial^2 S (E)/ \partial E_i \partial E_j)|_{E = E (a)} \|_{i, j = 0}^{d}$
by deleting the rows $i \not \in I$ and the columns $j \not \in I$, is negative. 
Denote this matrix as $S_{I}'' (a)$. 
\end{dfn}

Assume that our thermodynamic system $\Lambda$ is linearly stable. 
Then, in particular, $S_{[d' + 1, d]}'' (a) < 0$, $a \in \Lambda$, 
where 
\begin{equation*}
[d' + 1, d] := \lbrace d' + 1, d' + 2, \dots, d \rbrace. 
\end{equation*}
In quasithermodynamics the random vector $\delta E_{[d' + 1, d]}$ is taken to be Gaussian with 
the joint density distribution function 
\begin{multline}
f_{\delta E_{[d' + 1, d]}} (x; a) = 
\frac{(\det [- S_{[d' + 1, d]}'' (a)])^{1/ 2}}{(2 \pi k_B)^{(d - d')/ 2}} 
\times \\ \times
\exp \Big\lbrace 
- \frac{1}{2 k_B} \sum_{i, j = 1}^{d - d'} x_i [- S_{[d' + 1, d]}'' (a)]_{i, j} x_j
\Big\rbrace, 
\label{eq:density_f_E}
\end{multline}
where $x = (x_1, x_2, \dots, x_{d - d'}) \in \mathbb{R}^{d - d'}$, and $a \in M \subset \Lambda$. 

\vspace{0.2 true cm}
\noindent
\emph{Remark.} 
Note that by introducing a new physical constant $k_B$ into the theory (the Boltzmann constant) we actually 
make a step outside the paradigm of phenomenological thermodynamics. 
\hfill $\Diamond$
\vspace{0.2 true cm}

It is quite remarkable, that (according to Einstein) the associated \emph{intensive} quantities 
$\beta_{[d' + d]} := (\beta_{d' + 1}, \beta_{d' + 2}, \dots, \beta_{d})$
fluctuate as well. 
The equilibrium values in this case are $\beta_{[d' + 1, d]} (a) = (0, 0, \dots, 0)$, $a \in M \subset \Lambda$, 
and the fluctuations $\delta \beta_{[d' + 1, d]} := (\delta \beta_{d' + 1}, \delta \beta_{d' + 2}, \dots, \delta \beta_{d})$
form a Gaussian random vector with a joint density distribution function 
\begin{multline} 
f_{\delta \beta_{[d' + 1, d]}} (y) := 
\frac{1}{(2 \pi k_B)^{(d - d')/ 2} (\det [- S_{[d' + 1, d]}'' (a)])^{1/ 2}} 
\times \\ \times 
\exp \Big\lbrace 
- \frac{1}{2 k_{B}} \sum_{i, j = 1}^{d - d'} y_i ([- S_{[d' + 1, d]}'' (a)]^{-1})_{i, j} y_j
\Big\rbrace, 
\label{eq:density_f_beta}
\end{multline}
where $y = (y_1, y_2, \dots, y_{d - d'}) \in \mathbb{R}^{d - d'}$, and $a \in M \subset \Lambda$.

The classical probability theory (the axiomatics of A. N. Kolmogorov) identifies random variables 
with measurable functions on a space of events. 
More precisely, there is a probability model 
$(\Omega, \mathcal{F}, P)$, where $\Omega$ is a set of elementary events, $\mathcal{F}$ is a $\sigma$-algebra of events on $\Omega$, 
and $P$ is a probability measure on $(\Omega, \mathcal{F})$. 
The fluctuations $\delta E_{j}$ and $\delta \beta_{j}$, $j = d' + 1, d' + 2, \dots, d$, 
are measurable functions 
$\delta E_{j} : (\Omega, \mathcal{F}) \to (\mathbb{R}, \mathcal{B} (\mathbb{R}))$ and 
$\delta \beta_{j} : (\Omega, \mathcal{F}) \to (\mathbb{R}, \mathcal{B} (\mathbb{R}))$, 
where $\mathcal{B} (\mathbb{R})$ is the Borel $\sigma$-algebra on the real line $\mathbb{R}$.

Why would we actually assume that the fluctuations $\delta E_{j}$, $\delta \beta_j$, $j \in [d' + 1, d]$ 
should be modelled on a single probability space $(\Omega, \mathcal{F}, P)$? 
In quantum mechanics, if we consider the fluctuations of coordinates and momenta, this is not even possible 
(invoke Bell's inequalities \cite{Bell} and the Kochen-Specker type configurations (for an example, see \cite{Ruuge})). 
There we have a statistical operator $\wh{\rho} = \wh{\rho}^{\dagger} \geqslant 0$, $\mathit{Tr} \wh{\rho} = 1$, acting on a 
Hilbert space $\mathcal{H}$, in place of a probability measure $P$ on $(\Omega, \mathcal{F})$.  
In other words: 
\begin{itemize}
\item[] 
\emph{
It is natural to expect that if the thermodynamic system becomes smaller and smaller, 
then the probability model describing the fluctuations of extensive and intensive quantities becomes more and more quantum.}
\end{itemize}

More precisely, we expect that we can attach a Hilbert space $\mathcal{H} (a)$ to every point $a \in M \subset \Lambda$, 
and that the fluctuations $\delta E_{i}$ and $\delta \beta_j$ can be  
represented by self-adjoint operators $\wh{Q}_{i}$ and $\wh{P}_{j}$, respectively, $i, j \in [d' + 1, d]$. 
In the present paper I investigate the most natural possibility to define these operators. 

\begin{thm} 
Let $\Lambda \subset \mathbb{R}^{2 (d + 1)}$ be an abstract linearly stable thermodynamic system with an 
entropy function $S = S (E)$, where $E = (E_0, E_1, \dots, E_{d})$ are the extensive coordinates. 
Let $M \subset \Lambda$ be the submanifold corresponding to a reduction of degrees of freedom 
associated with $\lbrace E_{j} \rbrace_{j \in [d' + 1, d]}$, where $d' < d$. 
Put $\mathcal{H} (a) := L^2 (\mathbb{R}^{d - d'})$, for every $a \in M$. 
Then there exists $\psi (x; a) \in \mathcal{H} (a)$, where $x = (x_1, x_2, \dots, x_{d - d'})$ varies over $\mathbb{R}^{d - d'}$, 
such that 
\begin{equation*} 
f_{\delta E_{[d' + 1, d]}} (x; a) = |\psi (x; a)|^2, \quad 
f_{\delta \beta_{[d' + 1, d]}} (y; a) = |\wt{\psi} (y; a)|^2, 
\end{equation*}
where $\wt{\psi} (y; a)$, $y = (y_1, y_2, \dots, y_{d - d'}) \in \mathbb{R}^{d - d'}$, 
is the $(2 k_B)$-Fourier transform of $\psi (x; a)$ with respect to $x$. 
\end{thm}

\noindent
\emph{Proof.} 
Recall, that $h$-Fourier transform of $\varphi (x) \in L^{2} (\mathbb{R}^{n})$ is defined as follows: 
\begin{equation*} 
\wt{\varphi} (y) := (2 \pi h)^{n/ 2} \int_{\mathbb{R}^{n}} d x\, e^{- i y x/ h} \varphi (x), 
\end{equation*}
where $h > 0$, $y \in \mathbb{R}^{n}$, $n \in \mathbb{Z}_{> 0}$. 
If $A = \| A_{i, j} \|_{i, j = 1}^{n} > 0$ is a constant matrix, then for 
\begin{equation*} 
\varphi_{h} (x; A) := \frac{2^{n/ 4} (\det A)^{1/ 4}}{(2 \pi h)^{n/ 4}} 
\exp \Big\lbrace - \frac{1}{2 h} \sum_{i, j = 1}^{n} x_i A_{i, j} x_j \Big\rbrace, 
\end{equation*}
where $x = (x_1, x_2, \dots, x_n) \in \mathbb{R}$, we have: 
$\int_{\mathbb{R}^{n}} d x\, |\varphi_{h} (x)|^{2} = 1$, and the $h$-Fourier transform 
of $\varphi_{h} (x; A)$ with respect to $x$ is of the shape
$\wt{\varphi}_{h} (y; A) = \varphi_{h} (y; A^{-1})$. 
Take $a \in M \subset \Lambda$ and substitute: 
\begin{equation*} 
 n = d - d', \quad 
h = 2 k_{B}, \quad 
A = - S_{[d' + 1, d]}'' (a). 
\end{equation*}
It is straightforward to check that 
\begin{equation*} 
\psi (x; a) := \varphi_{2 k_B} (x; - S_{[d' + 1, d]}'' (a))
\end{equation*}
realizes the statement of the theorem. \qed 
\vspace{0.2 true cm}

\noindent 
\emph{Remark.} 
If we perceive the function constructed in the theorem in analogy with a coherent state in quantum mechanics, 
then we make a deep philosophical ``discovery'': 
the Planck constant $\hbar$ corresponds to the Boltzmann constant $k_B$ multiplied by two. 
\hfill $\Diamond$ 
\vspace{0.2 true cm} 

Do other thermodynamic wavefunctions $\psi (x; a) \in L^{2} (\mathbb{R}_{x}^{d - d'})$, $a \in M \subset \Lambda$, 
actually make sense or is it just a fancy property of Gaussian exponents? 
At least it is natural to expect that the functions ``similar'' to $\varphi_{2 k_B} (x; - S_{[d' + 1, d]}'' (a))$ \emph{do} make sense 
and can describe a deviation of the state of the system from the thermodynamic equilibrium. 
Take $x^{0} \in \mathbb{R}^{d - d'}$ and $y^0 \in \mathbb{R}^{d - d'}$ and consider a \emph{complex} thermodynamic ``wavefunction'' 
\begin{equation*} 
\psi_{x^{0}, y^{0}} (x; a) := e^{i y^{0} x/ (2 k_B)} \varphi_{2 k_B} (x - x^{0}; - S_{[d' + 1, d]}'' (a)). 
\end{equation*}
Note that the corresponding Weyl-Wigner function ($2 k_B$ in place of $\hbar$) is a Gaussian exponent 
concentrated in a point $(x^{0}, y^{0}) \in \mathbb{R}^{2 (d - d')} (x, y)$. 
Can thermodynamic ``wavefunctions'' be complex? 
Put 
\begin{equation*} 
\wh{Q}_j := x, \quad 
\wh{P}_{j} := - i (2 k_B) \frac{\partial}{\partial x_j}, 
\end{equation*}
for $j = 1, 2, \dots, d - d'$ (self-adjoint operators on $L^2 (\mathbb{R}^{d - d'} (x))$ 
corresponding to multiplication and derivation). 
Then for the moments of the fluctuations $\delta E_{d' + j}$ and $\delta \beta_{d' + j}$ 
in a state of thermodynamic equilibrium $a \in M \subset \Lambda$ we have: 
\begin{equation*} 
\begin{gathered}
\langle (\delta E_{d' + j})^{m} \rangle = 
\int_{\mathbb{R}^{d - d'}} dx \, \psi_{0, 0}^{*} (x; a), \wh{Q}_{j}^{m} \psi_{0, 0} (x; a), \\
\langle (\delta \beta_{d' + j})^{m} \rangle = 
\int_{\mathbb{R}^{d - d'}} dx \, \psi_{0, 0}^{*} (x; a), \wh{P}_{j}^{m} \psi_{0, 0} (x; a), 
\end{gathered}
\end{equation*}
where $m \in \mathbb{Z}_{> 0}$, $j = 1, 2, \dots, d - d'$, and the star denotes the complex conjugation. 
In particular: $\langle \delta E_{d' + j} \rangle = 0$ and $\langle \delta \beta_{d' + j} \rangle = 0$, $j = 1, 2, \dots, d - d'$. 
If we replace $\psi_{0, o} (x; a)$ with $\psi_{x^{0}, y^{0}} (x; a)$, then we obtain: 
\begin{equation*} 
\langle \delta E_{d' + j} \rangle = x_{j}^{0}, \quad 
\langle \delta \beta_{d' + j} \rangle = y_{j}^{0}, 
\end{equation*}
where $j = 1, 2, \dots, d - d'$. 

Consider a model example. Let $\Lambda = \Lambda^{(0)} \times \Lambda^{(1)}$ 
be a thermodynamic system consisting of two subsystems with entropy functions 
$S^{(\alpha)} = S^{(\alpha)} (U^{(\alpha)}, V^{(\alpha)}, N^{(\alpha)})$, $\alpha = 0, 1$ 
(the arguments are internal energy, volume, and number of moles). 
Put $(E_0, E_1, \dots, E_{5}) = 
(U^{(0)} + U^{(1)}, V^{(0)}, N^{(0)}, V^{(1)}, N^{(1)}, (U^{(0)} - U^{(1)})/ 2)$, and 
reduce with respect the coordinate $E_{5}$. 
We have $d' = 4$. 
The systems are in thermal contact. 
Consider a discrete analogue of the process of the exchange of energy. 
Suppose that it takes place in ``quanta'' and in discrete steps in time: 
a subsystem $\alpha = 0$ can release a fixed amount of energy $u > 0$ and the system $\alpha = 1$ then absorbs it, 
or vice versa: the system $\alpha = 1$ releases $u > 0$ and the system $\alpha = 0$ absorbs it. 
If there is a tendency that one of the cases takes place more often than the other then we obtain 
a non-zero value of $\langle \delta E_{5} \rangle$, which can be interpreted as an existence of a \emph{flow} of 
internal energy from one system to another. 
The intensive coordinate $\beta_5$ corresponding to $E_5 = (U^{(0)} - U^{(1)})/ 2$ is 
$\beta_{5} = \beta^{(0)} - \beta^{(1)}$, where $\beta^{(\alpha)}$ is the inverse absolute temperature in the subsystem $\alpha = 0, 1$. 
A tendency to observe more often that one of the quantities $\beta^{(0)}$, $\beta^{(1)}$, 
is greater than the other, 
leads to a non-zero value of $\langle \delta \beta_5 \rangle$. 
It can be interpreted as a gradient in inverse temperature, i.e. as an existence of a thermodynamic \emph{force}. 
In a state of thermodynamic equilibrium the thermodynamic forces and flows vanish. 
We conclude that a complex thermodynamic wavefunction 
$\psi_{x^{0}, y^{0}} (x; a)$
could describe a state near a thermodynamic equilibrium $a \in M \subset \Lambda$ 
with non-zero thermodynamic forces and flows. 

\vspace{0.2 true cm}
\noindent
\emph{Remark.} 
For the fluctuations $\delta \beta_j$, $\delta E_l$, $j, l \in [d' + 1, d]$, 
in the state $\psi_{0, 0} (-; a)$ holds:   
\begin{equation*} 
\int d x\, \psi_{0, 0}^{*} (x; a) ((\wh{P}_{j} \wh{Q}_{l} + \wh{Q}_{l} \wh{P}_{j})/ 2) \psi_{0, 0} (x; a) = 0, 
\end{equation*} 
where $j, l \in [d' + 1, d]$. 
Therefore, 
the linear correlation coefficient 
\begin{equation*}
\mathit{Corr} (\delta \beta_j, \delta E_l) = 0, 
\end{equation*}
for $j, l \in [d' + 1, d]$. 
This fact is compatible with the point of view of B. Mandelbrot \cite{Mandelbrot} on the 
fluctuations of intensive thermodynamic quantities 
(for a review see \cite{Ruppeiner}).  
Loosely speaking, one interprets 
the symbols $\delta \beta_j$, $j \in [d' + 1, d]$, 
as \emph{fluctuations of estimators} of parameters of a probability distribution related to $\delta E_l$, $l \in [d' + 1, d]$,  
based on statistical samples. 
In Landau-Lifshits \cite{LandauLifshits} one finds something completely different: 
``their'' fluctuations, which we denote $\Delta_{a} \beta_{j}$ and $\Delta_{a} E_{l}$, $j, l \in [d' + 1, d]$,  
$a \in M \subset \Lambda$, are 
linearly linked via the equations of state: 
\begin{equation} 
\Delta_{a} \beta_{j} = 
\sum_{l = d' + 1}^{d} \frac{\partial^2 S (E)}{\partial E_{j} \partial E_{l}} \Big|_{E = E (a)} \Delta_{a} E_{l}, 
\label{eq:LandauLifshits}
\end{equation}
where $S = S (E)$ is the entropy as a function of $E = (E_0, E_1, \dots, E_d)$, $j \in [d' + 1, d]$, 
and 
\begin{equation*}
\mathit{Corr} (\Delta_{a} \beta_{j}, \Delta_{a} E_{l}) = \delta_{j, l},  
\end{equation*}
where $\delta_{j, l}$ is the Kronecker delta, $j, l \in [d' + 1, d]$. 
We can now perceive this formula in analogy with semiclassical quantum mechanics, $\hbar \to 0$. 
Having a semiclassical wavefunction 
$\psi_{\hbar} (q) = \exp (i A (q)/ \hbar) \varphi_{\hbar} (q)$, $q \in \mathbb{R}^{n}$,  
where $A (q)$ is a \emph{real} smooth function (the classical action), 
and $\varphi_{\hbar} (q)$ is a complex smooth function, 
it is possible to write it as 
a superposition of coherent states concentrated in the points of a Lagrangian manifold 
$L \subset \mathbb{R}_{p, q}^{2 n}$, $p = (p_1, p_2, \dots, p_n)$, $q = (q_1, q_2, \dots, q_n)$, 
described by the equations: $p_j = \partial A (q)/ \partial q_j$, $j = 1, 2, \dots, n$. 
The Landau-Lifshits equations are just a linearised analogue of these equations of classical mechanics. 
A generic quasithermodynamic ``wavefunction'' is a superposition of 
coherent states concentrated in different points $(x^{0}, y^{0})$, where  
$x^{0} = (x_{1}^{0}, x_{2}^{0}, \dots, x_{d - d'}^{0})$, 
$y^{0} = (y_{1}^{0}, y_{2}^{0}, \dots, y_{d - d'}^{0})$, and 
$x_{l}^{0}$ and $y_{j}^{0}$ are values of $\Delta_a E_{d' + l}$ and $\Delta_a \beta_{d' + j}$, respectively, 
$j, l \in [1, d - d']$, linked by the linear equations above \eqref{eq:LandauLifshits}. 
A generic quasithermodynamic ``mixed state'' is a convex linear combination of orthogonal projectors 
corresponding to quasithermodynamic ``wavefunctions''. 
\hfill $\Diamond$ 

\vspace{0.2 true cm}

\noindent
\emph{Remark.}
In case of a distributed thermodynamic system 
localized in a space domain $\mathcal{D} \subset \mathbb{R}^{3}$ 
and on a time interval $\mathcal{T} \subset \mathbb{R}$, 
one should consider the space-time \emph{densities} of the fluctuations: 
$\delta \beta_j (t, \vec r)$ and $\delta E_l (t, \vec r)$, $j, l \in [d' + 1, d]$, 
$t \in \mathcal{T}$, 
$\vec r \in \mathcal{D}$. 
The quantities of interest are the correlation functions 
$\langle 
\xi_{j_0}^{(m_0)} (z_{0}) 
\xi_{j_1}^{(m_1)} (z_{1}) \dots \xi_{j_{n - 1}}^{(m_{n - 1})} (z_{n - 1})\rangle$, 
where 
$n \in \mathbb{Z}_{> 0}$, 
$z_{\alpha} \in \mathcal{T} \times \mathcal{D}$, 
$j_{\alpha} \in [d' + 1, d]$, 
$m_{\alpha} \in \lbrace 0, 1 \rbrace$, 
$\alpha \in [0, n - 1]$, 
the space-time 
points $\lbrace z_{\alpha} \rbrace_{\alpha = 0}^{n - 1}$ are mutually distinct, 
and $\xi_{l}^{(m)} = \delta \beta_{l}$, if $m = 0$, and 
$\xi_{l}^{(m)} = \delta E_{l}$, if $m = 1$, for $l \in [d' + 1, d]$.
It is natural to expect in analogy with the quantum field theory that 
these correlation functions can be expressed as 
$\int D \xi (\cdot) \exp ((2 k_B)^{-1} S[\xi (\cdot)])
\prod_{\alpha = 0}^{n - 1} \xi_{j_{\alpha}}^{(m_{\alpha})} (z_{\alpha})$, where 
$S [\xi (\cdot)]$ is a functional defined on 
$\xi (\cdot) = \lbrace \xi_{l}^{(m)} (\cdot) \rbrace_{l \in [d' + 1, d], m = 0, 1}$, and 
$\xi_{l}^{(m)} (\cdot)$ vary over smooth functions on $\mathcal{T} \times \mathcal{D}$.
\hfill $\Diamond$ 
\vspace{0.2 true cm}

\section{Tropical Pauli problem}

Speaking about a thermodynamic system $\Lambda$ such as a one component gas $S = S (U, V, N)$ 
(entropy $S$ as a function of internal energy $U$, volume $V$, and \emph{number of moles} $N$), 
we use the units of measurement which are adapted to the level of classical physics: 
length is measured, for instance, in centimeters, but not in angstroms or in parsecs, 
time is measured is seconds, but not in femtoseconds or billions of years. 
The number of moles $N$ in our system is measured in the scale of $1, 2, 3, \dots$, but we are 
not considering billionth fractions of moles or billions of moles. 
A convenient unit of measurement of absolute temperature on a classical level is one Kelvin. 

On the other hand the approximate value of the Boltzmann constant in the CGS system is 
\begin{equation*} 
k_B = 1.38 \times 10^{-16} \mathit{erg} K^{-1}. 
\end{equation*}
This numeric value is quite small and therefore the Gaussian exponents considered in the previous section 
are actually quite sharp. The standard deviations of the fluctuating quantities are proportional to $\sqrt{k_{B}}$, 
so the effect of the quasithermodynamic fluctuations is numerically in the $10^{- 8}$ scale. 
Therefore we may formally consider $k_B$ as a small parameter, just like $\hbar$ is considered small in 
semiclassical quantum mechanics. 

Let us restrict to the case where we have only \emph{one} reduced degree of freedom. 
In the notation of the previous section: $d' = d - 1$. 
For a fixed point $a \in \Lambda$, 
we may also adjust the units of measurement
of quasithermodynamic fluctuations $\delta E_d$ by taking a 
linear transformation $\xi^{(a)} = c (a) \delta E_d$, $c (a) \in \mathbb{R}_{> 0}$, 
in such a way that the corresponding probability density function is of the shape: 
\begin{equation*} 
f_{\xi^{(a)}} (x) = (2 \pi k_{B})^{- 1/ 2} \exp ( - x^2/ (2 k_B)), 
\end{equation*}
where $x \in \mathbb{R}$. Note that the physical dimension of the quantity $\xi^{(a)}$ is then $[\xi^{(a)}] = [k_B]^{1/ 2}$. 
Put $\eta^{(a)} := c (a)^{-1} \delta \beta_{d}$. Then the corresponding probability density is of the shape: 
\begin{equation*} 
f_{\eta^{(a)}} (y) = (2 \pi k_{B})^{- 1/ 2} \exp ( - y^2/ (2 k_B)), 
\end{equation*}
where $y \in \mathbb{R}$. Note that $[\eta^{(a)}] = [k_B]^{1/ 2}$ as well. 
The solution of the Pauli problem corresponding to the pair of functions $(f_{\xi^{(a)}}, f_{\eta^{(a)}})$ is of the shape: 
\begin{equation*} 
\varphi_{h} (x) = (\pi h)^{- 1/ 4} \exp (- x^2/ (2 h)), 
\end{equation*}
where one needs to substitute $h = 2 k_{B}$. 
We have: 
\begin{equation*} 
f_{\xi^{(a)}} (x) = |\varphi_{2 k_B} (x)|^2, \quad 
f_{\eta^{(a)}} (y) = |\wt{\varphi}_{2 k_B} (y)|^2, 
\end{equation*}
where $x, y \in \mathbb{R}$, and $\wt{\varphi}_{h}$ is the $h$-Fourier transform of $\varphi_{h}$. 

In quantum mechanics, if we take a semiclassical wavefunction $\psi_{\hbar} (x) = f (x) \exp (i S (x)/ \hbar)$, $x \in \mathbb{R}$, 
where $f$ and $S$ are smooth functions and $S$ is real, then the $\hbar \to 0$ asymptotics of its $\hbar$-Fourier transform 
is (under some natural conditions) also a fast oscillating exponent. 
Assume that the equation $y = \partial S (x)/ \partial x$ has a unique 
solution $x = \bar x (y)$, so that we can define the Legendre transform 
\begin{equation*} 
\wt{S} (y) = (S (x) - y x)|_{x = \bar x (y)}, 
\end{equation*}
in every point $y \in \mathbb{R}$. 
Then for the $\hbar$-Fourier transform we have $\wt{\psi}_{\hbar} (y) = \exp (i \wt{S} (y)/ \hbar) (g (y) + O (\hbar))$, 
where $g (y)$ is smooth. 
Let $x (0) = 0$ and write 
\begin{equation*} 
\partial S (x)/\partial x^2 = u (x), \quad 
\partial \wt{S} (y)/\partial y^2 = w (y), \quad 
\end{equation*}
There is a link between the derivatives of the functions $u$ and $w$: 
\begin{equation} 
\Big( \frac{\partial}{\partial y} \Big)^{n} w (y) \Big|_{y = 0} = 
- \Big( \frac{1}{u (x)} \frac{\partial}{\partial x} \Big)^{n} \frac{1}{u (x)} \Big|_{x = 0},  
\label{eq:link_uw}
\end{equation}
where $n \in \mathbb{Z}_{\geqslant 0}$. 

In tropical (older name -- idempotent) mathematics, an analogue of the 
Fourier transform is the Legendre transform (the latter can be written as an idempotent integral). 
The link 
\eqref{eq:link_uw}
mentioned 
was obtained for the real functions $u$ and $w$, but we can formally apply it 
on \emph{complex} functions $u (x)$ and $w (y)$. 

Observe that the quasithermodynamic ``wavefunction'' $\varphi_{2 k_B} (x)$ is of the form 
$\varphi_{2 k_B} (x) = (\pi \, 2 k_B)^{- 1/ 4} 
\exp (i S (x)/ (2 k_B))$, where $S (x) = i x^2/ 2$ is purely imaginary. 
For the $(2 k_B)$-Fourier transform we have: 
$\wt{\varphi}_{2 k_B} (y) = (\pi \, 2 k_B)^{- 1/ 4} 
\exp (i S (y)/ (2 k_B))$, $S (y) = i y^2/ 2$. 
Having this in mind, 
it is natural to define a \emph{tropical} analogue of the Pauli problem as follows: 
given a pair of smooth \emph{real} functions $U (x)$ and $W (y)$ defined 
in some neighbourhoods of $x = 0$ and $y = 0$, respectively, 
find a pair of complex functions $u (x)$ and $w (y)$, such that 
\begin{equation*} 
\mathrm{Im} (u (x)) = U (x), \qquad 
\mathrm{Im} (w (y)) = W (y), 
\end{equation*}
for which the formula 
\eqref{eq:link_uw} 
linking the derivatives in $x = 0$ and $y = 0$ holds. 

\vspace{0.2 true cm} 
\noindent 
\emph{Example 4.} 
A pair of constant functions $u (x) = i$ and $w (y) = i$ is a solution of the tropical Pauli problem 
corresponding to $U (x) = 1$ and $W (y) = 1$. 
\hfill $\Diamond$
\vspace{0.2 true cm}

The insight is that one should reconstruct the functions not from the real, but from the imaginary 
parts of the function and its ``Legendre transform''. 
In the introduction we have mentioned a somewhat counter intuitive fact about the Pauli problem: 
if it has a solution then there can be in fact several solutions. 
Let us look at what happens in the tropical case. 
Let 
\begin{equation*} 
u (x) = i \Big( 1 + \sum_{m = 0}^{\infty} \frac{x^m}{m!} [u_m + i \lambda_m] \Big), \quad 
w (y) = i \Big( 1 + \sum_{m = 0}^{\infty} \frac{y^m}{m!} [w_m + i \rho_m] \Big), 
\end{equation*} 
where the real coefficients $u_m$ and $w_m$ are given, and the real coefficients $\lambda_m$ and $\rho_m$ 
need to be reconstructed, $m = 1, 2, \dots$. 
The condition on the derivatives yields: 
\begin{equation*} 
i^n [w_n + i \rho_n] = \Big( \frac{1}{i^{-1} u (x)} \frac{\partial}{\partial x} \Big)^{n} \frac{1}{i^{-1} u (x)} \Big|_{x = 0}, 
\end{equation*}
where $n \in \mathbb{Z}_{> 0}$, 
so the coefficients $\rho_{n}$, $n = 1, 2, \dots$, become immediately known once we know $\lambda_{m}$, $m = 1, 2, \dots$. 
The first four equalities $n = 1, 2, 3, 4$ yield: 
\begin{equation*} 
\begin{aligned}
- w_1 &= \mathrm{Im} \lbrace u_1 + i \lambda_1 \rbrace, \\
w_2 &= \mathrm{Re} \lbrace 
(u_2 + i \lambda_2) - 3 (u_1 + i \lambda_1)^2
\rbrace, \\
w_3 &= \mathrm{Im} \lbrace 
(u_3 + i \lambda_3) - 10 (u_1 + i \lambda_1) (u_2 + i \lambda_2) + 15 (u_1 + i \lambda_1)^3
\rbrace, 
\end{aligned}
\end{equation*}
and 
\begin{multline*} 
- w_4 = \mathrm{Re} \lbrace 
(u_4 + i \lambda_4) - 15 (u_1 + i \lambda_1) (u_3 + i \lambda_3) - 10 (u_2 + i \lambda_2)^2 
+ \\ + 
105 (u_1 + i \lambda_1)^2 (u_2 + i \lambda_2) - 
105 (u_1 + i \lambda_1)^4
\rbrace
\end{multline*}
From the first equation we find $\lambda_1 = - w_1$, and substituting it in the second equation we obtain a condition: 
\begin{equation} 
w_2 = u_2 - 3 \mathrm{Re} \lbrace (u_1 - i w_1)^2 \rbrace. 
\label{eq:condition_w2u2}
\end{equation}
From the third equation, taking into account $\lambda_1 = - w_1$, we express $\lambda_3$ as a linear function of $\lambda_2$: 
\begin{equation*} 
\lambda_3 = 10 u_1 \lambda_2 + w_3 - 10 w_1 u_2 - 15 \mathrm{Im} \lbrace (u_1 - i w_1)^3 \rbrace. 
\end{equation*}
Substituting this into the fourth equation, we obtain a \emph{quadratic} equation on $\lambda_2$: 
\begin{equation} 
\lambda_2^2 + 6 u_1 w_1 \lambda_2 + q = 0, 
\label{eq:quadratic_eqn_Dq}
\end{equation}
where 
\begin{multline*} 
q := - \frac{3 w_1}{2} \lbrace w_3 - 10 w_1 u_2 - 15 \mathrm{Im} [(u_1 - i w_1)^3] \rbrace 
+ 
\frac{1}{10} \lbrace 
w_4 + u_4 - \\ - 15 u_1 u_3 - 10 u_2^2 + 105 (u_1^2 - w_1^2) u_2 - 105 \mathrm{Re} [(u_1 - i w_1)^4]
\rbrace. 
\end{multline*}
Since $\lambda_2$ \emph{should} be real, there is an important condition on the discriminant: 
\begin{equation} 
D := (6 u_1 w_1)^2 - 4 q \geqslant 0. 
\label{eq:discriminant}
\end{equation}
It is of interest to notice, that if we express $\rho_2$ via $\lambda_1$ and $\lambda_2$, then we obtain: 
$\lambda_2 + (- \rho_2) = - 6 u_1 w_1$, i.e. $- \rho_2$ is just the second root of this quadratic equation.  
Denote the two roots mentioned $\lambda_{2}^{(i)}$, $i = 1, 2$. 

Look now at the equations corresponding to higher degrees $n = 5, 6, \dots$. 
If $f (x)$ is a smooth function such that $f (x) \not = 0$, 
then one can prove by induction ($n \geqslant 5$) the following formula: 
\begin{multline*} 
\Big( \frac{1}{f (x)} \frac{\partial}{\partial x} \Big)^{n} \frac{1}{f (x)} = 
(- 1) \frac{f^{(n)}}{f^{n + 2}} 
+ C_{n + 2}^{2} \frac{f' f^{(n - 1)}}{f^{n + 3}} 
+ \\ + C_{n + 2}^{3} \frac{f'' f^{(n - 2)}}{f^{n + 3}} 
- \mathit{Tt}_n \frac{(f')^2 f^{(n - 2)}}{f^{n + 4}} + 
 \frac{Q_{n} (f, f' , \dots, f^{(n - 3)})}{f^{2 n + 1}}, 
\end{multline*}
where on the right-hand side we omit the argument $x$ in the function $f$ and its derivatives, 
$Q_n$ is a polynomial in $n - 2$ variables, 
$C_{n + 2}^{2} = (n + 1) (n + 2)/ 2$ and $C_{n + 2}^{3} = n (n + 1) (n + 2)/ 6$ are binomial coefficients, 
and $\mathit{Tt}_{n} := C_{C_{n + 2}^{2}}^{2} = n (n + 1) (n + 2) (n + 3)/ 8$ 
are the so-called \emph{tritriangular numbers}, the generating function: 
\begin{equation*} 
\frac{3 x}{(1 - x)^5} = \sum_{m = 1}^{\infty} \mathit{Tt}_{m} x^{m}. 
\end{equation*}
We have: 
\begin{equation*} 
\begin{aligned}
(- 1)^{k} w_{2 k + 1} &= \mathrm{Im} \Big\lbrace 
\Big( 
\frac{1}{i^{-1} u (x)} \frac{\partial}{\partial x}  
\Big)^{2 k + 1} \frac{1}{i^{-1} u (x)}
\Big\rbrace \Big|_{x = 0}, \\
(- 1)^{k + 1} w_{2 k + 2} &= 
\mathrm{Re} \Big\lbrace 
\Big( 
\frac{1}{i^{-1} u (x)} \frac{\partial}{\partial x}  
\Big)^{2 k + 1} \frac{1}{i^{-1} u (x)}
\Big\rbrace \Big|_{x = 0}, 
\end{aligned}
\end{equation*}
so if $k \geqslant 2$, we obtain from the first equation: 
\begin{equation*} 
(- 1)^{k} w_{2 k + 1} = \mathrm{Im} \lbrace 
- [u_{2 k + 1} + i \lambda_{2 k + 1}] + 
C_{2 k + 3}^{2} [u_1 + i \lambda_1] (u_{2 k} + i \lambda_{2 k}) + Z_{2 k + 1}
\rbrace, 
\end{equation*}
where $Z_{2 k + 1} = Z_{2 k + 1} (\lambda_1, \lambda_2, \dots, \lambda_{2 k - 1})$.
This yields a linear link between $\lambda_{2 k + 1}$ and $\lambda_{2 k}$: 
\begin{equation*} 
\lambda_{2 k + 1} - C_{2 k + 3}^{2} u_1 \lambda_{2 k} + 
\lbrace (-1)^{k} w_{2 k + 1} - C_{2 k + 3}^{2} \lambda_{1} u_{2 k} + \mathrm{Im} (Z_{2 k + 1})\rbrace = 0. 
\end{equation*}
On the other hand, the second equation yields: 
\begin{multline*} 
(- 1)^{k + 1} w_{2 k + 2} = \\ = 
\mathrm{Re} \lbrace 
- [u_{2 k + 2} + i \lambda_{2 k + 2}] 
+ C_{2 k + 4}^{2} [u_1 + i \lambda_1] (u_{2 k + 1} + i \lambda_{2 k + 1})
 + \\ +  
[C_{2 k + 4}^{3} (u_2 + i \lambda_2) - \mathit{Tt}_{2 k + 2} (u_1 + i \lambda_1)^2] (u_{2 k} + i \lambda_{2 k}) + \wt{Z}_{2k + 2}
\rbrace, 
\end{multline*}
where $\wt{Z}_{2 k + 2} = \wt{Z}_{2 k + 2} (\lambda_1, \lambda_2, \dots, \lambda_{2 k - 1})$. 
From here, invoking the linear link between $\lambda_{2 k + 1}$ and $\lambda_{2 k}$, we obtain: 
\begin{equation*} 
\lambda_{2 k} \lbrace \lambda_1 u_1 [C_{2 k + 4}^{2} C_{2 k + 3}^{2} - 2 \mathit{Tt}_{2 k + 2}] + C_{2 k + 4}^{3} \lambda_2 \rbrace + \wh{Z}_{2k + 1} = 0, 
\end{equation*}
where $\wh{Z}_{2 k + 1} = \wh{Z}_{2 k + 1} (\lambda_1, \lambda_2, \dots, \lambda_{2 k - 1})$. 
If the expression in the curly brackets in the last equation is not zero, then all the 
coefficients $\lambda_{2 k}$, and therefore $\lambda_{2 k + 1}$, $k \geqslant 2$, are determined. 
It remains to notice that: 
\begin{equation*}
\begin{gathered}
C_{2 k + 4}^{2} C_{2 k + 3}^{2} - 2 \mathit{Tt}_{2 k + 2} = 
- 2 (k + 1) (k + 2) (2 k + 3), \\
C_{2 k + 4}^{3} = 2 (k + 1) (k + 2) (2 k + 3)/ 3, 
\end{gathered}
\end{equation*}
so we even obtain a condition $- 3 \lambda_1 u_1 + \lambda_2 \not = 0$ which does not depend on the parameter $k \geqslant 2$. 
Taking into account that $\lambda_1 = - w_1$ and that $\lambda_2$ satisfies the quadratic equation $\lambda_2^2 + 6 u_1 w_1 \lambda_2 + q = 0$, 
we reduce the condition just to $q \not = (3 u_1 w_1)^2$. 

Assume the conditions 
\eqref{eq:discriminant}, \eqref{eq:condition_w2u2}, 
and $q \not = (3 u_1 w_1)^2$, 
are satisfied. 
We have the power series $1 + \sum_{m = 1}^{\infty} [u_m + i \lambda_m] x^m/ m!$ and 
$1 + \sum_{m = 1}^{\infty} [w_m + i \rho_m] y^m/ m!$. 
To avoid a question about the \emph{convergence} of these power series, it is natural to consider a \emph{truncated} tropical Pauli problem. 
Let $n_0 \in \mathbb{Z}_{\geqslant 0}$. 
Let $U, W \in C^{n_0} (\mathbb{R})$ be a pair of real-valued functions 
having a continuous derivative of order $n_0$ in a neighbourhood of zero. 
Then the aim is to construct a pair of polynomials $u (x)$ and $w (y)$ over $\mathbb{C}$ of degree at most $n_0$, 
such that 
\begin{equation*}
\mathrm{Im} (u (x)) = U (x) + O (x^{n_0 + 1}), \quad 
\mathrm{Im} (w (y)) = W (y) + O (y^{n_0 + 1}), 
\end{equation*} 
and  the formula 
\eqref{eq:link_uw}
induced by the Legendre transform linking the derivatives of order $n$ holds for $n \leqslant n_0$. \
We term $n_0$ the \emph{degree of truncation}. 

\begin{prop} 
Let $U (x)$ and $W (y)$ be a pair of smooth real functions such that 
$U (x) = 1 + \sum_{m = 1}^{\infty} u_m x^m/ m!$ and $W (y) = 1 + \sum_{m = 1}^{\infty} w_m y^m/ m!$. 
If the coefficients $\lbrace u_m \rbrace_{m = 1}^{4}$ and $\lbrace w_m \rbrace_{m = 1}^{4}$ 
satisfy the conditions: 
$w_2 = u_2 - 3 \mathrm{Re} \lbrace (u_1 - i w_1)^2 \rbrace$, $D > 0$, and $q \not = (3 u_1 w_1)^2$, 
where $q$ and $D$ are the constant term and the discriminant of the quadratic equation 
\eqref{eq:quadratic_eqn_Dq}, 
respectively, 
then for any degree of truncation $n_0 \in \mathbb{Z}_{\geqslant 2}$, the truncated tropical Pauli problem 
has exactly two solutions. In the limit $D \to +0$, the two solutions coincide. 
\end{prop}

\vspace{0.2 true cm} 
\noindent 
\emph{Proof.} 
See the explanations above. \qed 
\vspace{0.2 true cm}

One may try to ``improve'' the formulae \eqref{eq:density_f_E}, \eqref{eq:density_f_beta}, for 
the densities of distributions $f_{\delta E_d} (x)$ and $f_{\delta \beta_d} (y)$ as follows. 
For $a \in M \subset \Lambda$ put $S_{\geqslant 2} (x; a) := S (E_{[0, d - 1]} (a), E_d (a) + x) - 
\lbrace S (E (a)) + x \partial S (E)/ \partial E_d |_{E = E (a)} \rbrace$, and 
$\Phi_{\geqslant 2} (y; a) := S_{\lbrace d \rbrace} (E_{[0, d - 1]} (a), \beta_d (a) + y) - 
\lbrace 
S_{\lbrace d \rbrace} (E_{[0, d - 1]} (a), \beta_d (a)) + y \partial S_{\lbrace d \rbrace} (E_{[0, d - 1]} (a), \beta_d)/ \partial \beta_d |_{\beta_d = \beta_d (a)}
\rbrace$, 
where $S_{\lbrace d \rbrace} (E_{[0, d - 1]}, \beta_d)$ is the Legendre transform of the entropy function $S = S (E)$ with respect 
to the last argument, $E_{[0, d - 1]} = (E_0, E_1, \dots, E_{d - 1})$, 
$S_{\lbrace d \rbrace} (E_{[0, d - 1]} (a), \beta_d (a)) = S (E (a)) - \beta_d (a) E_d (a)$, $a \in \Lambda$ 
(we assume that this transform is defined). 
Replace $f_{\delta E_d} (x)$ and $f_{\delta \beta_d} (y)$ with 
\begin{equation*} 
\begin{gathered}
\wt{f}_{\delta E_d} (x) = Z (a)^{-1} \exp (S_{\geqslant 2} (x; a)/ k_B), \\
\wt{f}_{\delta \beta_d} (y) = \wt{Z} (a)^{-1} \exp (- \Phi_{\geqslant 2} (y; a)/ k_B), 
\end{gathered}
\end{equation*}
where $Z (a)$ and $\wt{Z} (a)$ are the normalizing factors. 
Note that in the point $(x, y) = (0, 0)$ we have 
$\partial \Phi_{\geqslant 2} (y; a)/ \partial y^2 |_{y = 0} \, \partial S_{\geqslant 2} (x; a)/ \partial x^2 |_{x = 0} = -1$, 
and that is why we need a minus sign in the second exponent since $\partial S_{\geqslant 2} (x; a)/ \partial x^2 < 0$, for $x \not = 0$. 
More generally: 
\begin{equation*} 
\Big( \frac{\partial}{\partial y} \Big)^{n + 2} \Phi_{\geqslant 2} (y; a) \Big|_{y = 0} = 
- \Big( \frac{1}{S_{\geqslant 2}'' (x; a)} 
\frac{\partial}{\partial x} \Big)^n \frac{1}{S_{\geqslant 2}'' (x; a)} \Big|_{x = 0}, 
\end{equation*}
where $S_{\geqslant 2}'' (x; a) := \partial^2 S_{\geqslant 2} (x; a)/ \partial x^2$, and $n = 0, 1, \dots$. 
We may assume that the coordinates $E_d$ and $\beta_d$ are chosen in such a way that 
\begin{equation*} 
\partial^2 S_{\geqslant 2} (x; a)/ \partial x^2 = U (x), \quad 
\partial^2 \Phi_{\geqslant 2}(y; a)/ \partial y^2 = W (y), 
\end{equation*}
where the functions $U (x)$ and $W (y)$ are like in the proposition 2. 
Taking $n = 1$ and $n = 2$ we obtain: 
$w_1 = u_1$ and $w_2 = - u_2 + 3 u_1^2$. 
On the other hand, in the proposition 2 
we have a condition 
$w_2 - u_2 + 3 [u_1^2 - w_1^2] = 0$, so in the end this yields: 
\begin{equation*}
w_2 = u_2, \quad 
2 u_2 = 3 u_1^2.  
\end{equation*}
This is a very strong condition on $S = S (E)$ and there is no special reason for it to hold. 
Therefore typically one can not span a quasithermodynamic $2 k_B \to 0$ ``wavefunction'' over 
$\wt{f}_{\delta E_d} (x)$ and $\wt{f}_{\delta \beta_d} (y)$, and 
such fluctuations deviate from a pure state.

\section{Subtropical Pauli problem}

One may consider other improvements in the formulae for $f_{\delta E_d} (x)$ and $f_{\delta \beta_d} (y)$ 
corresponding to the fluctuations of the reduced degree of freedom $(\beta_{d}, E_{d})$ 
in a linearly stable thermodynamic system $\Lambda \subset \mathbb{R}^{2 (d + 1)} (\beta, E)$. 
We keep the notation $M \subset \Lambda$ for the submanifold corresponding to $\beta_d = 0$. 
It is perhaps more natural to look for a thermodynamic ``wavefunction'' $\varphi_{2 k_B} (x)$ of the form: 
\begin{equation*}
\varphi_{h} (x) = \exp (- x^2/ (2 h)) u_{h} (x), 
\end{equation*}
where $h = 2 k_B$, and $u_h (x)$ is smooth and expands in asymptotic power series in $h \to 0$, $x \in \mathbb{R}$. 
Assuming that $u_h (x)$ does not grow too fast at infinity, we have a formula: 
\begin{equation*} 
\exp \Big( \frac{h}{2} \frac{\partial^2}{\partial z^2} \Big) u_h (z) = 
(2 \pi h)^{1/ 2} \int_{\mathbb{R}} d x \, 
\exp (- (x - z)^2/ (2 h)) u_{h} (x), 
\end{equation*}
where $z \in \mathbb{R}$. 
Then for the $h$-Fourier transform $\wt{\varphi}_{h} (y)$ we obtain: 
\begin{equation*} 
\wt{\varphi}_{h} (y) = \exp (- y^2/ (2 h)) w_{h} (x), 
\end{equation*}
where $w_h (y)$ expands into an asymptotic power series in $h \to 0$, 
\begin{equation*} 
w_{h} (y) \sim \exp \Big( \frac{h}{2} \frac{\partial^2}{\partial z^2} \Big) u_h (z) \Big|_{z = - i y}, 
\end{equation*}
where $y \in \mathbb{R}$. 
Let $u_{h} (x) = \exp (\wh{A}_{h} (x))$ and $w_{h} (y) = \exp (\wh{B}_{h} (y))$, where 
\begin{equation*} 
\wh{A}_{h} (x) \sim \sum_{m = 0}^{\infty} \frac{h^m}{m!} [A_m (x) + i f_m (x)], \,\,
\wh{B}_{h} (y) \sim \sum_{m = 0}^{\infty} \frac{h^m}{m!} [B_m (y) + i g_m (y)], 
\end{equation*}
where $A_m$, $f_m$, $B_m$, and $g_m$, are real-valued functions, $m = 0, 1, \dots$. 
If these functions are analytical, 
\begin{equation*} 
\begin{gathered}
A_m (x) = \sum_{n = 0}^{\infty} \frac{A_{m, n} x^n}{n!}, \quad 
f_m (x) = \sum_{n = 0}^{\infty} \frac{f_{m, n} x^n}{n!}, \\
B_m (y) = \sum_{n = 0}^{\infty} \frac{B_{m, n} y^n}{n!}, \quad 
g_m (y) = \sum_{n = 0}^{\infty} \frac{g_{m, n} y^n}{n!}, 
\end{gathered}
\end{equation*}
then we can write: 
\begin{equation} 
\exp(\wh{B}_{h} (i x)) \sim \exp \Big( \frac{h}{2} \frac{\partial^2}{\partial x^2} \Big) \exp(\wh{A}_h (x)), 
\label{eq:link_Ahat_Bhat}
\end{equation}
where $x \in \mathbb{R}$. This formula determines a link between the coefficients $\lbrace A_{m, n} + i f_{m, n} \rbrace_{m, n = 0}^{\infty}$ and 
$\lbrace B_{m, n} + i g_{m, n} \rbrace_{m, n = 0}^{\infty}$, and it suffices to assume that $x$ varies in a small neighbourhood of zero. 

In the Pauli problem one assumes the knowledge of $|\varphi_{h} (x)|^2$ and $|\wt{\varphi}_{h} (y)|^2$. 
It follows, that one can mimic it as follows: 
given the collections of coefficients $\lbrace A_{m, n} \rbrace_{m, n = 0}^{\infty}$ and $\lbrace B_{m, n} \rbrace_{m, n = 0}^{\infty}$, 
reconstruct the collections $\lbrace f_{m, n} \rbrace_{m, n = 0}^{\infty}$ and $\lbrace g_{m, n} \rbrace_{m, n = 0}^{\infty}$. 
More precisely, 
given a pair of real-valued functions $A^h (x)$ and $B^h (y)$ admitting an asymptotic power series expansion in $h$ with analytic coefficients, 
construct a pair of real-valued functions $f^h (x)$ and $g^h (y)$ which also admit asymptotic power series expansions in $h$ with analytic coefficients, 
such that the formula 
\eqref{eq:link_Ahat_Bhat}
holds for $\wh{A}_h (x) := A^h (x) + i f^h (x)$, $\wh{B}_h (y) := B^h (y) + i g^h (y)$. 

We refer to the problem above as a \emph{subtropical Pauli problem} 
(the small parameter $h = 2 k_B$ describes a ``deformation'' of the tropical case). 
It is natural to consider also a \emph{truncated} subtropical Pauli problem 
in order to avoid the questions about convergence of the corresponding power series.  
Fix parameters $m_0, n_0 \in \mathbb{Z}_{\geqslant 0}$ (the \emph{degrees of truncation}) and require that 
\begin{equation} 
\Big( \frac{\partial}{\partial x} \Big)^{n} \Big\lbrace 
\exp(\wh{B}_{h} (i x)) - \exp \Big( \frac{h}{2} \frac{\partial^2}{\partial x^2} \Big) \exp(\wh{A}_h (x)) 
\Big\rbrace \Big|_{x = 0} \sim O (h^{m_0 + 1}), 
\label{eq:link_Ahat_Bhat_truncated}
\end{equation}
for every $n = 0, 1, \dots, n_0$. 
Basically it means that we consider the subtropical Pauli problem on polynomials.
Note that without loss of generality we can also assume that $f_{m, 0} = 0$ and $g_{m, 0} = 0$, for all $m = 0, 1, \dots, m_0$, 
since the ``wavefunctions'' $\varphi_{h} (x)$ and $\wt{\varphi}_{h} (y)$, $h = 2 k_B$, are defined up to a constant phase factor. 
One can also truncate only with respect to $h$. 
In this case we require that a solution $(f^{h}, g^{h})$ is given by the polynomials in $h$ of degree $n \leqslant n_0$,

Let us look at what happens with the coefficients $f_{m, n}$ and $g_{m, n}$, $m, n = 0, 1, \dots$, in more detail. 
Commuting the exponents on the right-hand side and then taking a logarithm in the equation 
\eqref{eq:link_Ahat_Bhat_truncated}
yields: 
\begin{equation*} 
\wh{B}_{h} (i x) \sim \wh{A}_{h} (x) + 
\log \Big\lbrace 
1 + \sum_{m = 1}^{\infty} \frac{1}{m!} \Big( \frac{h}{2} \Big)^{m} 
\Big[ \Big(\frac{\partial \wh{A}_h (x)}{\partial x} + \frac{\partial}{\partial x} \Big)^{2 m} 1 \Big] 
\Big\rbrace. 
\end{equation*}
Expanding the logarithm into a Taylor power series and then regrouping the terms, yields: 
\begin{multline} 
\wh{B}_{h} (i x) \sim \wh{A}_{h} (x) + \sum_{n = 1}^{\infty} \frac{1}{n!} \Big( \frac{h}{2} \Big)^{n} 
\sum_{p = 1}^{n} \frac{(- 1)^{p + 1}}{p} 
\times \\ \times 
\sum_{\substack{m_1, m_2, \dots, m_p = 1, \\ m_1 + m_2 + \dots + m_p = n}}^{n - p + 1} 
\frac{n!}{m_1! m_2! \dots m_p!} 
\prod_{\alpha = 1}^{p} \Big[ 
\Big( \frac{\partial \wh{A}_h (x)}{\partial x} + \frac{\partial}{\partial x} \Big)^{2 m_{\alpha}} 1
\Big]. 
\label{eq:A_B_Q_series}
\end{multline}
Introduce now a notation. Let $\Phi (x)$ be a power series in $x$ with real coefficients. 
Put:
\begin{equation*}
 \Phi_{\mathit{even}} (x) := (\Phi (x) + \Phi (- x))/ 2, \quad  
\Phi_{\mathit{odd}} (x) := (\Phi (x) - \Phi (- x))/ 2. 
\end{equation*}
Observe that if $x$ is real, then 
$\Phi_{\mathit{even}} (i x)$ is real, while $\Phi_{\mathit{odd}} (i x)$ is purely imaginary. 
Separating the real and imaginary parts, we obtain: 
\begin{equation*} 
\begin{gathered} 
B_{\mathit{even}}^{h} (i x) + i g_{\mathit{odd}}^{h} (i x) \sim 
A^{h} (x) + \mathrm{Re} ( \wh{Q}_{h}^{(f)} (x) ), \\ 
- i B_{\mathit{odd}}^{h} (i x) + g_{\mathit{even}}^{h} (i x) \sim 
f^{h} (x) + \mathrm{Im} ( \wh{Q}_{h}^{(f)} (x) ), 
\end{gathered}
\end{equation*}
where $\wh{Q}_{h}^{(f)} (x)$ is the series on the right-hand side in \eqref{eq:A_B_Q_series}, 
$\wh{B}_{h} (i x) \sim \wh{A}_{h} (x) + \wh{Q}_{h}^{(f)} (x)$. 
Since the series $\wh{Q}_{h}^{(f)} (x)$ is expressed via $f^h (x)$, we can conclude that 
once $f^{h} (x)$ is defined, we immediately know $g^{h} (x)$. 
Projecting on the even and odd parts, we conclude: 
\begin{equation} 
\begin{aligned} 
B_{\mathit{even}}^{h} (i x) &\sim A_{\mathit{even}}^{h} (x) + 
[\mathrm{Re} ( \wh{Q}_{h}^{(f)} (x) )]_{\mathit{even}}, \\
- i B_{\mathit{odd}}^{h} (i x) &\sim f_{\mathit{odd}}^{h} (x) + 
[\mathrm{Im} ( \wh{Q}_{h}^{(f)} (x) )]_{\mathit{odd}}. 
\end{aligned}
\label{eq:B_A_Re_Im_Q}
\end{equation}
The coefficients at $h^{0}$ yield:
\begin{equation*} 
(B_{0})_{\mathit{even}} (i x) = (A_0)_{\mathit{even}} (x), \quad 
- i (B_{0})_{\mathit{odd}} (i x) = (f_0)_{\mathit{odd}} (x). 
\end{equation*}
The first equality is a condition on the input data, and the second equality determines the odd part of $f_0 (x)$. 

Look at the coefficients at $h^1$. 
The first equation in \eqref{eq:B_A_Re_Im_Q} together with 
$(f_0)_{\mathit{odd}} (x) = - i (B_{0})_{\mathit{odd}} (i x)$ yields: 
\begin{equation*} 
\frac{1}{2} \Big( \frac{\partial (f_{0})_{\mathit{even}} (x)}{\partial x} \Big)^{2} = c (x), 
\end{equation*}
where 
\begin{multline} 
c (x) := - (B_{1})_{\mathit{even}} (i x) + (A_{1})_{\mathit{even}} (x) 
+ \\ + 
\frac{1}{2} 
\Big[ \Big( \frac{\partial A_0 (x)}{\partial x} \Big)^2 + \frac{\partial^2 A_0 (x)}{\partial x^2} \Big]_{\mathit{even}} 
- \frac{1}{2} \Big[ \frac{\partial (- i (B_{0})_{\mathit{odd}} (i x))}{\partial x} \Big]^2. 
\label{eq:function_c}
\end{multline} 
Note that $c (x)$ must vanish in $x = 0$ since 
$(\partial (f_0)_{\mathit{even}} (x)/ \partial x)|_{x = 0} = 0$. 
This can be perceived as a condition on $(B_{1})_{\mathit{even}} (0) - (A_{1})_{\mathit{even}} (0)$.  
Furthermore, 
there is an important condition: $c (x) \geqslant 0$. 
Since without loss of generality $f^{h} (x)|_{x = 0} = 0$, the function $f_{0} (x)$ is now known (if $c (x) \not = 0$ then there are two branches). 
The coefficients at $h^1$ in the second equation 
in \eqref{eq:B_A_Re_Im_Q}
determine $(f_1)_{odd} (x)$: 
\begin{equation*} 
- i (B_{1})_{\mathit{odd}} (i x) = (f_1)_{\mathit{odd}} (x) + 
\frac{1}{2} 
\mathrm{Im} \Big[ \Big( \frac{\partial [A_0 + i f_0] (x)}{\partial x} + \frac{\partial}{\partial x} \Big)^2 1 \Big]_{\mathit{odd}}. 
\end{equation*}

Look now at the coefficients at $h^{k + 1}$ for $k \geqslant 1$. 
The second equation 
in \eqref{eq:B_A_Re_Im_Q}
determines $(f_{k + 1})_{\mathit{odd}} (x)$ since the coefficient 
in $[\mathrm{Im} ( \wh{Q}_{h}^{(f)} (x) )]_{\mathit{odd}}$ at $h^{k + 1}$ depends only on $\lbrace f_{l} (x) \rbrace_{l \leqslant k}$. 
The analysis of the first equation 
in \eqref{eq:B_A_Re_Im_Q}
is slightly more complicated. 
In the definition of $\wh{Q}_{h}^{(f)} (x)$ we have a sum over $n = 1, 2, \dots$. 
Let us extract the term $n = 1$ explicitly: 
\begin{equation*} 
\wh{Q}_{h}^{(f)} (x) = \frac{h}{2} \Big[ \Big( 
\frac{\partial \wh{A}_{h} (x)}{\partial x} + \frac{\partial}{\partial x}
\Big)^2 1 \Big] + \wt{Q}_{h}^{(f)} (x),  
\end{equation*}
where $\wt{Q}_{h}^{(f)} (x) \sim O (h^2)$. 
Then we obtain: 
\begin{multline*} 
\frac{(B_{k + 1})_{\mathit{even}} (i x) - (A_{k + 1})_{\mathit{even}} (x)}{(k + 1)!} = 
\frac{1}{2 k!} \mathrm{Re} \Big\lbrace
\frac{\partial^2 [A_{k} + i f_{k}] (x)}{\partial x^2} 
+ \\ +
\sum_{l = 0}^{k} 
\frac{k!}{l! (k - l)!} 
\frac{\partial [A_{l} + i f_{l}] (x)}{\partial x}
\frac{\partial [A_{k - l} + i f_{k - l}] (x)}{\partial x} 
\Big\rbrace_{\mathit{even}} 
+ \\ + (\mathrm{Re} (\wt{Q}_{h}^{(f)} (x)))_{\mathit{even}}. 
\end{multline*}
Therefore we obtain a linear equation on $\partial (f_{k})_{\mathit{even}} (x)/ \partial x$: 
\begin{equation}
\frac{\partial (f_{0})_{\mathit{even}} (x)}{\partial x}
\frac{\partial (f_{k})_{\mathit{even}} (x)}{\partial x} + 
\frac{\partial (f_{0})_{\mathit{odd}} (x)}{\partial x}
\frac{\partial (f_{k})_{\mathit{odd}} (x)}{\partial x} + 
Z_{k + 1}^{(f)} (x) = 0, 
\label{eq:f0_fk_Z}
\end{equation}
where $Z_{k + 1}^{(f)} (x)$ requires a knowledge only of $\lbrace f_{l} (x) \rbrace_{l \leqslant k - 1}$. 
The left-hand side of this equality must vanish in $x = 0$, what is basically a condition on 
$(B_{k + 1})_{\mathit{even}} (0) - (A_{k + 1})_{\mathit{even}} (0)$, $k \geqslant 1$. 
If $c'' (x)|_{x = 0} \not =  0$, then $(\partial^2 (f_0)_{\mathit{even}} (x)/ \partial x^2) |_{x = 0} \not = 0$, 
and differentiating the left-hand side of \eqref{eq:f0_fk_Z} many times, one can 
find recursively all the derivatives of $(f_{k})_{\mathit{even}} (x)$ in $x = 0$.  
Without loss of generality, $(f_k)_{\mathit{even}} (x)|_{x = 0} = 0$. 
\begin{thm} 
Assume that $(B_{0})_{\mathit{even}} (i x) = (A_0)_{\mathit{even}} (x)$, and that 
the function $c (x)$ defined in \eqref{eq:function_c} 
is non-negative in a neighbourhood of $x = 0$, and $c'' (0) \not = 0$. 
Then for any degrees of truncation $(m_0, n_0) \in (\mathbb{Z}_{> 0})^2$ 
there exists a polynomial $P (h)$ in $h = 2 k_{B}$ of degree $m_0$ with constant real coefficients, 
such that the truncated subtropical Pauli problem for 
$(A^{h} + h P (h), B^{h})$ 
has exactly two 
solutions vanishing in zero 
(polynomials in $h = 2 k_B$ and $x$ of degrees at most $m_0$ and $n_0$, respectively).
  
\end{thm}

\noindent
\emph{Proof.} 
See the explanations above. 
The polynomial $P (h)$ is necessary to adjust the values of 
$(B_k)_{\mathit{even}} (0) - (A_{k})_{\mathit{even}} (0)$, $k = 1, 2, \dots$. 
\qed 
\vspace{0.2 true cm} 

\noindent
\emph{Example 5.} 
Fix a point $a \in M \subset \Lambda$, and 
consider a thermodynamic ``wavefunction'' of the shape: 
\begin{equation*} 
\varphi_{h} (x; a) := (\pi h)^{- 1/ 4} 
\exp (- (1 + i h \sigma) x^2/ (2 h)) 
\end{equation*}
where $h = 2 k_B$, and $\sigma = \sigma (a)$ is a real parameter. 
We have: $\varphi_{h} (x; a) = (\pi h)^{- 1/ 4} \exp (- x^2/ (2 h)) \exp (\wh{A}_h (x))$, where 
$\wh{A}_{h} (x) = - i \sigma x^2/ 2$. 
Then 
$\wt{\varphi} (y; a) = (\pi h)^{- 1/ 4} \exp (- y^2/ (2 h)) \exp (\wh{B}_{h} (y))$ 
(the $h$-Fourier transform of $\varphi_{h} (- ; a)$), where 
\begin{equation*} 
\wh{B}_{h} (y) = - \frac{1}{2} \log (1 + i h \sigma) + 
\frac{y^2}{2} \frac{i \sigma}{1 + i h \sigma}. 
\end{equation*}
This yields the following data 
for a truncated  subtropical Pauli problem: 
\begin{equation*} 
A_0 (x) = 0, \quad 
A_1 (x) = 0, \quad 
B_0 (y) = 0, \quad 
B_1 (y) = \sigma^2 y^2/ 2. 
\end{equation*}
For the quantity $c (x)$ we obtain: $c (x) = \sigma^2 x^2/ 2$, so 
$\partial f_0 (x)/ \partial x = \pm \sigma x$. Assuming that $f_0 (0) = 0$, we obtain: 
$f_0 (x) = \pm \sigma x^2/ 2$. 
Put 
\begin{equation*} 
\varphi_{h}^{(\alpha)} (x; a) := (\pi h)^{- 1/ 4} 
\exp \Big(- \frac{x^2}{2 h} [1 + (-1)^{\alpha} i h \sigma (a)] \Big)
\end{equation*}
where $\alpha = 0, 1$, 
Look at the 2-dimensional subspace $L = L (a)$ in $\mathcal{H} = L^2 (\mathbb{R})$
spanned over $\varphi_{h}^{(0)} (-; a)$ and $\varphi_{h}^{(1)} (-; a)$. 
Let us 
leave out the fixed point $a \in M \subset \Lambda$ in the notation and write just 
$\varphi^{(\alpha)} (x)$ in place of $\varphi^{(\alpha)} (x; a)$. 
The inner products between the basis vectors $\varphi_{h}^{(\alpha)} (x)$, $\alpha = 0, 1$, are of the shape: 
\begin{equation*} 
(\varphi_{h}^{(0)}, \varphi_{h}^{(0)}) = 1, \quad 
(\varphi_{h}^{(1)}, \varphi_{h}^{(1)}) = 1, \quad 
(\varphi_{h}^{(0)}, \varphi_{h}^{(1)}) = (1 - i h \sigma)^{- 1/ 2}, 
\end{equation*}
where one should take the branch 
$(1 - i h \sigma)^{- 1/ 2} \to 1$, as $h \sigma \to 0$, of the square root. 
Consider an orthonormal basis in $L = L (a)$: 
\begin{equation*} 
\psi_{h}^{(0)} (x) := \varphi_{h}^{(0)} (x), \quad 
\psi_{h}^{(1)} (x) := \frac{\varphi_{h}^{(1)} (x) - \varphi_{h}^{(0)} (x) (\varphi_{h}^{(0)}, \varphi_{h}^{(1)})}{
(1 - |(\varphi_{h}^{(0)}, \varphi_{h}^{(1)})|^2)^{1/ 2}},  
\end{equation*}
where we leave out the argument $a$ in $\psi_{h}^{(0)}$, $\psi_{h}^{(1)}$, as well as in $\varphi_{h}^{(0)}$, $\varphi_{h}^{(1)}$. 
A generic statistical operator $\wh{\rho}$ concentrated on $L = L (a)$ is (in the Dirac notation) of the shape: 
\begin{equation*} 
\wh{\rho} = \sum_{m, n = 0, 1}^{}p_{m, n} | \psi_{h}^{(m)} \rangle \langle \psi_{h}^{(n)} |, 
\end{equation*}
where 
the coefficients $p_{m, n} = p_{m, n} (a) \in \mathbb{C}$ form a $2 \times 2$ matrix 
$P = \| p_{m, n} \|_{m, n = 0, 1}$ such that $P = P^{\dagger} \geqslant 0$ and $\mathrm{Tr} (P) = 1$. 
These additional data provided by the matrix $P = P (a)$ define a ``deformation'' of the quasithermodynamic fluctuation theory 
in the point $a \in M \subset \Lambda$. 
The variances $\langle (\delta E_d)^{2} \rangle$ and $\langle (\delta \beta_d)^{2} \rangle$ are computed as 
\begin{equation*} 
\begin{gathered}
\langle (\delta E_d)^{2} \rangle = \sum_{m, n = 0, 1} p_{m, n} \langle \psi_{2 k_B}^{(n)} | \wh{Q}^{2} | \psi_{2 k_B}^{(m)} \rangle, \\
\langle (\delta \beta_d)^{2} \rangle = \sum_{m, n = 0, 1} p_{m, n} \langle \psi_{2 k_B}^{(n)} | \wh{P}^{2} | \psi_{2 k_B}^{(m)} \rangle, 
\end{gathered}
\end{equation*}
where $\wh{Q} = x$ (multiplication by $x$), and $\wh{P} = - i (2 k_B) \partial/ \partial x$. 
The covariance coefficient is computed as   
\begin{equation*} 
\langle \delta \beta_d \delta E_d \rangle = \sum_{m, n = 0, 1} p_{m, n} \langle \psi_{2 k_B}^{(n)} | \wh{R} | \psi_{2 k_B}^{(m)} \rangle, 
\end{equation*}
where $\wh{R} := (\wh{P} \wh{Q} + \wh{Q} \wh{P})/ 2$. 
If the statistical operator $\wh{\rho} = | \psi_{2 k_B}^{(0)} \rangle \langle  \psi_{2 k_B}^{(0)} |$, 
then we recover in the leading term in $2 k_B \to 0$ the formulae of the standard quasithermodynamics. 
\hfill $\Diamond$
\vspace{0.2 true cm}

Note that the case $c (x) = 0$ is actually admissible, but it is a little technical and we do not go into its details. 
A generalization of the tropical and subtropical Pauli problems to many dimensions is straightforward, but 
their analysis involves some overdetermined linear systems of equations. 
In principle, the general picture where an existing solution can have an ``antipode'' remains valid.

\section{Conclusions and discussion} 
The main motive of the present paper is that 
the \emph{probability model} describing quasithermodynamic fluctuations becomes more and more 
``\emph{quantum}'' if the thermodynamic system 
gets smaller and smaller. 
By that it is not implied that the Planck constant $\hbar$ starts to play a more visible role, 
but rather that the very ``nature'' of the probability model becomes different: 
\begin{itemize} 
\item[]
\emph{There is no particular reason to perceive 
the fluctuations of thermodynamic quantities 
as random variables sharing the same probability space.}
\end{itemize}
It is natural to expect an analogy with quantum mechanics, i.e. that 
at some point one needs a statistical operator on a Hilbert space to describe these fluctuations. 
A more detailed analysis implemented in the present paper 
shows that the role of $\hbar$ can be 
taken over by the Boltzmann constant $k_B$ multiplied by two.

We look at an abstract thermodynamic system $\Lambda$ and 
reduce it to $\Lambda'$ with respect to several degrees of freedom $n = \dim (\Lambda) - \dim (\Lambda')$. 
This leaves us a ``heritage'' consisting of  
fluctuating intensive and extensive thermodynamic quantities associated with these degrees of freedom, 
for which we state an analogue of the Pauli problem: 
reconstruct a ``wavefunction'' (more generally -- span a class of statistical operators) 
consistent with experimentally available data 
(the marginal probability densities of the collections of quantities which can be measured simultaneously).

Many important features of the thermodynamic Pauli problem can already be seen if $n = 1$ and in the 
quasithermodynamic (``semiclassical'') limit $2 k_B \to 0$. 
We formalize these observations in a theorem about a truncated subtropical Pauli problem 
proven in the main text. 
An important conclusion here is the following. 
In principle, to describe the fluctuations of thermodynamic quantities in analogy with quantum mechanics 
we need to attach a Hilbert space $\mathcal{H} (a) \simeq L^2 (\mathbb{R}^{n})$ to every point $a \in \Lambda'$ 
and look at statistical operators on this space. 
It turns out that if we are only interested in the first corrections to the standard 
quasithermodynamic fluctuation theory, 
then one can significantly simplify things 
considering instead of the infinite dimensional Hilbert space $\mathcal{H} (a)$ 
a 2-dimensional subspace $L (a) \subset \mathcal{H} (a)$. 
Typically (i.e. under the conditions of the theorem mentioned) the truncated subtropical Pauli problem 
is going to have \emph{two} solutions which correspond to a pair of quasithermodynamic ``wavefunctions'' 
$\varphi_{2 k_B}^{(0)} (x; a)$ and $\varphi_{2 k_B}^{(1)} (x; a)$, $x \in \mathbb{R}$. 
Knowing one of the functions allows to reconstruct the second one explicitlly. 
The space $L (a)$ is then 
\begin{equation*} 
L (a) := \mathrm{span}_{\mathbb{C}} \lbrace \varphi_{2 k_{B}}^{(0)} (-; a), \varphi_{2 k_B}^{(1)} (-; a) \rbrace. 
\end{equation*}
Instead of a generic statistical operator on $\mathcal{H} (a)$ one may approximate the fluctuations 
by a statistical operator $\wh{\rho} (a)$ concentrated on $L (a)$ 
(the operator $\wh{\rho} (a)$ in the latter case is, loosely speaking, nothing more but a $2 \times 2$ matrix). 
In the leading degree in particular cases one recovers the standard quasithermodynamic fluctuation theory, 
but the general formulae contain new \emph{interference terms}, which are of interest, for example, in 
the quantum information science.


\begin{thebibliography}{99} 



\bibitem{BalianValentin}
Balian, R.; Valentin, P.: 
Hamiltonian structure of thermodynamics with gauge. 
\emph{Eur. Phys. J. B Condens. Matter Phys.} \textbf{21} (2001), no. 2, 269--282


\bibitem{Kazinski}
Kazinski P. O.: 
Stochastic deformation of a thermodynamic symplectic structure. 
\emph{Phys. Rev. E} \textbf{79}, 011105 (2009) [9 pages] 


\bibitem{Lavenda}
Lavenda, B. H.: 
Thermodynamic uncertainty relations and irreversibility. 
\emph{Internat. J. Theoret. Phys.} \textbf{26} (1987), no. 11, 1069--1084


\bibitem{Mehrafarin}
Mehrafarin, M.: 
Canonical operator formulation of non-equilibrium thermodynamics. 
\emph{J. Phys. A: Math. Gen.} \textbf{16} (1993), 5351--5363


\bibitem{RudoiSukhanov}
Rudoi, Yu. G.; Sukhanov, A. D.: 
Thermodynamic fluctuations within the Gibbs and Einstein approaches. 
\emph{Phys. Usp.} \textbf{43} (2000), no. 12, 1169--1199



\bibitem{UffinkLith}
Uffink, J.; van Lith, J.: 
Thermodynamic uncertainty relations. 
\emph{Found. Phys.} \textbf{29} (1999), no. 5, 655--692


\bibitem{VelazquezCurilef}
Velazquez, L.; Curilef, S.: 
A thermodynamic fluctuation relation for temperature and energy. 
\emph{J. Phys. A: Math. Theor.} \textbf{42} (2009) 095006 (19pp)




\bibitem{OnsagerMachlup} 
Onsager, L.; Machlup, S.: 
Fluctuations and irreversible processes. 
\emph{Phys. Rev.} \textbf{91} (1953), no. 6, 1505--1512

\bibitem{MachlupOnsager} 
Machlup, S.; Onsager, L.: 
Fluctuations and irreversible processes. II. Systems with kinetic energy. 
\emph{Phys. Rev.} \textbf{91} (1953), no. 6, 1512--1515

\bibitem{AcostaFernandez-de-CordobaIsidroSantander}
Acosta, D.; Fern\'andez de C\'ordoba, P.; Isidro, J.M.; Santander, J.L.G.: 
Emergent quantum mechanics as a classical, irreversible thermodynamics. 
\emph{Int. J. Geom. Methods Mod. Phys.} \textbf{10} (2013), no. 4, 1350007 [20 pages] 

\bibitem{Fernandez-de-CordobaIsidroPerea}
Fern\'andez de C\'ordoba, P.; Isidro, J.M.; Perea, M.H.: 
Emergent quantum mechanics as a thermal ensemble. 
\texttt{arXiv:1304.6295 [math-ph]} 




\bibitem{IbortMankoMarmocSimonicVentrigliac} 
Ibort, A.; Man'ko, V.I.; Marmoc, G.; Simonic, A.; Ventrigliac, F.: 
An introduction to the tomographic picture of quantum mechanics. 
\emph{Phys. Scr.} \textbf{79} (2009), 065013




\bibitem{Maslov_thermo1}
Maslov, V. P.:
Geometric quantization of thermodynamics, phase transitions and asymptotics at critical points. (Russian) 
\emph{Mat. Zametki} \textbf{56} (1994), no. 3, 155--156; 
translation in \emph{Math. Notes} \textbf{56} (1994), no. 3-4, 984--985 (1995)


\bibitem{Maslov_thermo2}
Maslov, V. P.: 
Analytic extension of asymptotic formulas, and the axiomatics of thermodynamics and quasithermodynamics.  (Russian)
\emph{Funktsional. Anal. i Prilozhen.}  \textbf{28}  (1994),  no. 4, 28--41, 95;  
translation in  \emph{Funct. Anal. Appl.}  \textbf{28}  (1994),  no. 4, 247--256 (1995)


\bibitem{Maslov_thermo3}
Maslov, V. P.: 
Geometric ``quantization'' of thermodynamics, and statistical corrections at critical points. (Russian) 
\emph{Teoret. Mat. Fiz.} \textbf{101} (1994), no. 3, 433--441; 
translation in \emph{Theoret. and Math. Phys.} \textbf{101} (1994), no. 3, 1466--1472 (1995) 


\bibitem{Maslov_ultra}
Maslov, V. P.: 
Ultrasecond quantization and ``ghosts'' in quantized entropy. (Russian)  
\emph{Teoret. Mat. Fiz.}  \textbf{129}  (2001),  no. 3, 464--490;  
translation in  \emph{Theoret. and Math. Phys.}  \textbf{129}  (2001),  no. 3, 1694--1716


\bibitem{Maslov_Nazaikinskii} 
Maslov, V. P.; Nazaikinskii, V. E.: 
The tunnel canonical operator in thermodynamics. (Russian)  
\emph{Funktsional. Anal. i Prilozhen.}  \textbf{40}  (2006),  no. 3, 12--29, 96;  
translation in  \emph{Funct. Anal. Appl.}  \textbf{40}  (2006),  no. 3, 173--187



\bibitem{Maslov_op_meth}
Maslov, V. P.: 
\emph{Operational methods.} 
Translated from the Russian by V. Golo, N. Kulman and G. Voropaeva. 
Mir Publishers, Moscow, 1976. 559 pp.



\bibitem{Maslov_asymp_meth}
Maslov, V. P.: 
\emph{Asymptotic methods and perturbation theory.}
``Nauka'', Moscow, 1988. 311 pp. ISBN: 5-02-013784-7 





\bibitem{Rovelli}
Rovelli, C.: 
Statistical mechanics of gravity and the thermodynamical origin of time. 
\emph{Class. Quantum Grav.} \textbf{10} (1993), 1549--1566


\bibitem{ConnesRovelli}
Connes, A.; Rovelli, C.: 
Von Neumann algebra automorphisms and time-thermodynamics relation in generally covariant quantum theories. 
\emph{Class. Quantum Grav.} \textbf{11} (1994), 2899--2917 


\bibitem{MontesinosRovelli}
Montesinos, M.; Rovelli, C.: 
Statistical mechanics of generally covariant quantum theories: a Boltzmann-like approach. 
\emph{Class. Quantum Grav.} \textbf{18} (2001), 555--569


\bibitem{Rajeev_contact} 
Rajeev, S. G.: 
Quantization of contact manifolds and thermodynamics.  
\emph{Ann. Physics}  \textbf{323}  (2008),  no. 3, 768--782.






\bibitem{AllahverdyanNieuwenhuizen}
Allahverdyan, A. E.;  Nieuwenhuizen, Th. M.: 
Explanation of the Gibbs paradox within the framework of quantum thermodynamics. 
\emph{Phys. Rev. E} \textbf{73}, (2006) 066119 [15 pages]

\bibitem{HenrichMichelMahler} 
Henrich, M. J.; Michel, M.; Mahler, G.: 
Small quantum networks operating as quantum thermodynamic machines. 
\emph{Europhys. Lett.} \textbf{76} (2006), no.6, 1057--1063 


\bibitem{QuanLiuSunNori}
Quan, H. T.;  Yu-xi Liu; Sun, C. P.; Nori, F.: 
Quantum thermodynamic cycles and quantum heat engines. 
\emph{Phys. Rev. E} \textbf{76}, (2007) 031105 [18 pages]


\bibitem{LindenPopescuShortWinter}
Linden, N.; Popescu, S.; Short, A. J.; Winter, A.: 
Quantum mechanical evolution towards thermal equilibrium. 
\emph{Phys. Rev. E} \textbf{79}, (2009) 061103 [12 pages]


\bibitem{SkrzypczykBrunnerLindenPopescu}
Skrzypczyk, P.; Brunner, N.; Linden, N.; Popescu, S.: 
The smallest refrigerators can reach maximal efficiency. 
\emph{J. Phys. A: Math. Theor.} \textbf{44} (2011) 492002 (7pp)


\bibitem{Maslov_Gibbs}
Maslov, V. P.: 
Mathematical resolution of the Gibbs paradox. (Russian). 
\emph{Mat. Zametki} \textbf{89} (2011), no. 2, 272--284; 
translation in \emph{Math. Notes} \textbf{89} (2011), no. 1--2, 266--276




\bibitem{Bell}
Bell, J. S.: 
On the problem of hidden variables in quantum mechanics.  
\emph{Rev. Modern Phys.}  \textbf{38} (1966), 447--452.



\bibitem{Ruuge}
Ruuge, A. E.: 
New examples of Kochen-Specker type configurations on three qubits. 
(submitted), \texttt{arXiv:1206.6999v1 [quant-ph]}


\bibitem{Mandelbrot} 
Mandelbrot, B.: 
The role of sufficiency and of estimation in thermodynamics. 
\emph{Ann. Math. Statist.} \textbf{33} (1962), 1021--1038


\bibitem{Ruppeiner}
Ruppeiner, G.: 
Riemannian geometry in thermodynamic fluctuation theory. 
\emph{Rev. Modern Phys.} \textbf{67} (1995), no. 3, 605--659.



\bibitem{LandauLifshits}
Landau, L. D.; Lifshitz, E. M.: 
\emph{Statistical Physics. Vol. 5 (3rd ed.)}. 
Butterworth-Heinemann, 1980.




\end{thebibliography}
\end{document}